%% file: allreduce.tex
\documentclass[journal,draftclsnofoot,onecolumn]{IEEEtran}
\IEEEoverridecommandlockouts
\usepackage{graphicx}
\usepackage{mathtools}
\usepackage{tikz}

\usetikzlibrary{decorations.markings}
\usetikzlibrary{arrows.meta}
\usetikzlibrary{positioning,fit,shapes.geometric}

\usepackage{amsthm}
\usepackage{amsmath}
\usepackage{amsfonts}
\usepackage{amssymb}
\usepackage{bm}

\usepackage[caption=false,font=footnotesize]{subfig}
\usetikzlibrary{arrows.meta}
\usepackage{enumitem}

\tikzset{
  ringcir/.style={circle, draw, fill=gray!20, minimum size=0.4cm, inner sep=2pt},
  ringlab/.style={midway, auto, swap, font=\small, inner sep=1.5pt},
  ringedge/.style={->, >={Stealth[scale=1.0]}, thick}
}

\newtheorem{theorem}{Theorem}
\newtheorem{lemma}[theorem]{Lemma}

\newtheorem{proposition}[theorem]{Proposition}
\theoremstyle{definition}
\newtheorem{definition}{Definition}
\newtheorem{construction}{Construction}
\newtheorem{example}{Example}
\theoremstyle{remark}
\newtheorem{remark}{Remark}

\usepackage{url}
\usepackage{ifthen}
\usepackage{cite}

\usepackage{xcolor}

\newcommand{\fullsum}[1]{%
    \tikz[baseline=(digit.base)]{%
        \draw[line width=0.4pt]
            (-0.55em,-0.55em) rectangle (0.55em,0.55em);
        \node[inner sep=0pt, outer sep=0pt] (digit) at (0,0) {#1};
    }%
}
\newcommand{\extra}[1]{\textcolor{blue}{\textbf{#1}}}

\begin{document}
\title{Linear Network Coding for Sum All-Reduce over Ring Networks} 

\author{Yi Chen, Zhuoqi Tu, Shenghao Yang%
\thanks{This paper was presented in part at the 2026 IEEE International Symposium on Information Theory (ISIT) \cite{tu2026linear}.}
\thanks{Y.~Chen and S.~Yang are with the School of Science and Engineering, The Chinese University of Hong Kong, Shenzhen, Guangdong 518172, China. Z.~Tu is with the Department of Information Engineering, The Chinese University of Hong Kong, Hong Kong SAR, China.}
\thanks{This work was supported in part by the National Natural Science Foundation of China under Grants 12141108 and 62171399, by the Shenzhen Fundamental Research Program under Grant JCYJ20241202124023031, and by the 1+1+1 CUHK-CUHK(SZ)-GDSTC Joint Collaboration Fund under Grants 2025A0505000079 and 2025A0505000082.}
}

\maketitle

\begin{abstract}
Sum All-Reduce is a fundamental collective communication primitive in which each node holds a local vector and every node must recover the componentwise sum of all vectors. In distributed training, this operation is often implemented over a ring topology using a reduce--multicast scheme. We study sum All-Reduce on directed rings via linear network coding, which includes reduce--multicast as a special case. We give a necessary and sufficient feasibility condition in terms of the coding matrices. For a ring with $N\geq 3$ nodes and unit-capacity links, where each node holds a vector containing $K$ symbols, we prove that the number of time units $T$ required by any feasible linear code satisfies $T\ge\max\bigl\{K+N-2,\ \lceil\frac{2(N-1)}{N}K\rceil\bigr\}$, giving a rate $K/T$ of at most $N/(2(N-1))$. The bound is tight over every field. When $K\ge N$, an interleaved reduce--multicast code achieves the bound; when $K<N$ and $K$ divides $N$, a reduce--partial-multicast code does. When $K<N$ and $K$ does not divide $N$, no coordinate-wise code---in particular, no reduce--multicast code---achieves the bound, while an explicit linear network code does.
\end{abstract}


\section{Introduction}
The rapid growth of artificial intelligence (AI) has driven the development of large-scale computing clusters containing tens of thousands of nodes, such as the GPU clusters used to train large language models (LLMs)~\cite{jiang2024megascale}. Efficient distributed computation on these clusters depends critically on \emph{collective communication}, including operations such as Reduce-Scatter, All-Gather, and All-Reduce~\cite{nccl}. In sum All-Reduce, each node holds a local vector of symbols, and every node must recover the componentwise sum of all local vectors. This operation is fundamental to distributed learning, where it is used to aggregate and synchronize gradients across nodes.

This paper studies sum All-Reduce over an $N$-node directed ring. Suppose that each directed link can transmit one field symbol per time unit. Each node initially holds a vector of $K$ source symbols, and every node must recover the corresponding vector of $K$ sum symbols. If this computation is completed within $T$ time units, we define the All-Reduce rate as $K/T$. The classical ring All-Reduce algorithm implements Reduce-Scatter followed by All-Gather with $K=N$. Conceptually, these operations contain two stages: a \emph{reduction} stage that combines the local vectors and a \emph{multicast} stage that distributes the result to every node. The classical ring All-Reduce delivers $N$ sum symbols to every node in $T=2(N-1)$ time units and therefore achieves the rate $\frac{N}{2(N-1)}$.

The bandwidth optimality of ring All-Reduce has previously been studied within the reduce--multicast framework~\cite{optimal_all_reduce}. For one sum symbol, a reduce--multicast implementation requires at least $N-1$ link transmissions to combine the $N$ source symbols and at least another $N-1$ link transmissions to deliver the resulting sum to all $N$ nodes. Thus, one symbol reduce--multicast task requires at least $2(N-1)$ link transmissions. If an algorithm processes the $K$ symbol coordinates independently, this accounting scales to at least $2K(N-1)$ transmissions. Because the directed ring contains $N$ unit-capacity links and can therefore carry at most $NT$ field symbols in $T$ time units, the resulting rate bound is $\frac{K}{T}\le\frac{N}{2(N-1)}$. This argument establishes the desired bound for algorithms whose communication can be decomposed into independent reduce--multicast tasks, one for each sum symbol.

Such a symbol-by-symbol decomposition, however, cannot
be assumed for a general jointly coded scheme. A transmitted symbol may depend on several source-symbol coordinates and therefore need not be attributable to any single sum symbol or any single reduce--multicast task. As a result, a transmission lower bound for one sum symbol cannot automatically be multiplied by $K$ to obtain a converse for schemes that jointly process all $K$ symbol coordinates. This distinction is well known in network function computation: an optimal single-instance scheme need not remain optimal when multiple instances are computed jointly, and coding across source symbols can improve the computation rate in some networks~\cite{Appuswamy11,Kowshik12,huang18}. A rigorous converse for coded All-Reduce must therefore analyze the joint transfer of all $K$ source symbols without presupposing a symbol-by-symbol decomposition.

Motivated by this gap, we formulate sum All-Reduce as a linear network-coding problem that contains reduce--multicast as a special case but also permits joint linear coding across coordinates. For $K$-symbol vectors on an $N$-node directed ring, we prove that every feasible linear code over an arbitrary field requires
\begin{equation}
\label{eq:intro-T-bound}
T\ge\max\left\{K+N-2,\ \left\lceil\frac{2(N-1)}{N}K\right\rceil\right\},
\end{equation}
so the rate is at most $\frac{N}{2(N-1)}$. The bound is tight for every $K$ and $N\ge 3$, but the achieving schemes depend on the relation between $K$ and $N$: reduce--multicast (or a simple variant) suffices in some regimes, while in others joint coding across coordinates is necessary.

From an information-theoretic perspective, sum All-Reduce is a distributed function-computation problem. Linear network coding provides a natural algebraic framework by allowing each transmitted symbol to be a linear combination of the locally available source symbols and previously received symbols~\cite{koetter2003algebraic,li2003linear}. The duality between multicast and sum computation explains the close relationship between the dissemination and reduction components of All-Reduce~\cite{koetter2004network}. Unlike ordinary multicast, however, no single node initially possesses the demanded sum, so standard multicast min-cut arguments do not directly characterize feasibility or rate.

Sum All-Reduce is also related to the \emph{multi-sum} or \emph{sum-network} problem~\cite{rai2013capacity,rai2012sum,ramamoorthy2013communicating}, where several sources hold independent symbols and several destinations require their sum. Prior work connects sum-network solvability to multiple-unicast network coding~\cite{rai2012sum,kamath2016two}, shows that pairwise connectivity need not suffice with three or more sources and destinations~\cite{ramamoorthy2013communicating}, and establishes that  linear solvability can be field-dependent~\cite{tripathy2015capacity}. Existing sum-network studies primarily concern directed acyclic networks over finite fields. All-Reduce, by contrast, typically involves networks with nontrivial cycles, a setting that remains largely unexplored. This paper addresses that gap by studying multi-sum over a directed ring: a $T$-time-unit execution is represented as a linear code on a directed acyclic time-expanded network, whose repeated structure yields algebraic feasibility conditions and a rate converse without assuming a symbol-by-symbol reduce--multicast decomposition. The directed ring is a natural starting point for cyclic topologies, since similar encoding, recoding, and decoding constraints appear along directed cycles in more general networks.

We formulate the problem over a general field $\mathbb F$, which may be finite or infinite. This setting includes the usual finite-field model of network coding and exact linear computation over fields such as $\mathbb R$. Recent work has studied floating-point implementations of algebraic network codes with controlled numerical error~\cite{tu2025sumreduction}. Our analysis concerns exact field operations. Numerical stability and finite-precision effects are left as separate implementation considerations.
Our main contributions are summarized as follows.
\begin{itemize}
    \item We formulate sum All-Reduce on an $N$-node directed ring as a linear network-coding problem on a directed acyclic time-expanded network. We derive a necessary and sufficient algebraic condition for the feasibility of a linear network code that computes the sum of $K$-symbol vectors in $T$ time units (Theorem~\ref{theorem:necsuf}).
    \item We prove a linear-coding converse for every $N\ge 3$ (Theorem~\ref{theorem:necessary}): every feasible linear code satisfies~\eqref{eq:intro-T-bound}. The first term, $T\geq K+N-2$, follows from a maximum-flow argument in the time-expanded network. The second term $T\geq \lceil\frac{2(N-1)}{N}K\rceil $ 
follows from the linear feasibility identities and bounds on
subspace dimensions.
    \item We show that given $N$ and $K$, the completion-time lower bound in \eqref{eq:intro-T-bound} is achievable over every field. When $K\ge N$, an interleaved reduce--multicast code meets the bound (Theorem~\ref{theorem:interleaved}). When $K<N$ and $K$ divides $N$, a reduce--partial-multicast code meets the bound. When $K<N$ and $K$ does not divide $N$, no coordinate-wise code---in particular no reduce--multicast code---meets the bound (Theorem~\ref{thm:k-less-n}), while an explicit linear network code does (Theorem~\ref{thm:construction}).
\end{itemize}

The remainder of the paper is organized as follows.
Section~\ref{sec:ring_all} introduces the directed-ring model and the All-Reduce problem.
Section~\ref{sec:linear_network_coding} develops the time-expanded network formulation and derives the necessary and sufficient feasibility conditions.
Section~\ref{sec:rate-upper-bounds} proves the linear-coding rate converse.
Section~\ref{sec:reduce-multicast-sufficiency} constructs codes that achieve the completion-time lower bound when $K\ge N$, and when $K<N$ and $K$ divides $N$, and proves that no coordinate-wise code meets the bound when $K$ does not divide $N$.
Section~\ref{sec:construct-network-coding} gives a linear network code that achieves the bound in the remaining case, over every field.
Finally, Section~\ref{sec:conclusion} discusses open problems and possible extensions to other network topologies.

\section{The All-Reduce Problem}
\label{sec:ring_all}

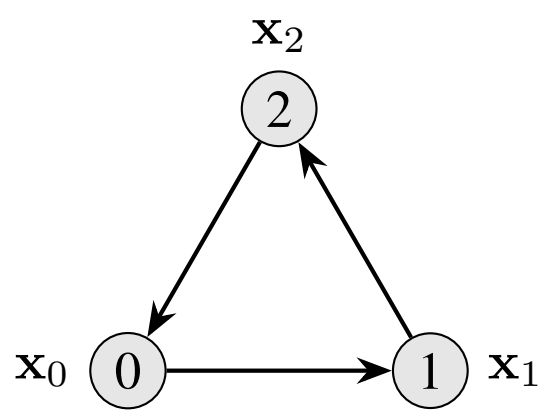
\begin{figure}[t]
    \centering
    \begin{tikzpicture}
        \node[ringcir,label=left:{$\mathbf x_0$}] (p1) at (-1,0) {0};
        \node[ringcir,label=right:{$\mathbf x_1$}] (p2) at (1,0) {1};
        \node[ringcir,label=above:{$\mathbf x_2$}] (p3) at (0,{sqrt(3)}) {2};
        \draw[ringedge] (p1) -- (p2);
        \draw[ringedge] (p2) -- (p3);
        \draw[ringedge] (p3) -- (p1);
    \end{tikzpicture}
    \caption{A directed $3$-node ring. Node $i$ holds a local message vector $\mathbf x_i\in\mathbb F^K$.}
    \label{fig:3-node-ring}
\end{figure}

Fix an integer \(N\ge 3\). We consider the directed ring
\(G=(\mathcal V,\mathcal E)\) with node set
\(\mathcal V=\{0,1,\ldots,N-1\}\) and edge set
\(\mathcal E=\{(i,(i+1)\bmod N):i\in\mathcal V\}\).
Fig.~\ref{fig:3-node-ring} illustrates the case \(N=3\).
Let \(\mathbb F\) be an arbitrary field, finite or infinite.
Each edge can transmit one symbol from \(\mathbb F\) per time unit, without error.

Let \(K\) be a positive integer. Each node \(i\in\mathcal V\) initially holds a message vector of $K$ symbols:
\[
\mathbf x_i=[x_{i,0},x_{i,1},\ldots,x_{i,K-1}]\in\mathbb F^K.
\]
The \emph{\(K\)-symbol sum All-Reduce problem} requires every node to
recover the componentwise sum
\(\sum_{i\in\mathcal V}\mathbf x_i\).
If this is accomplished in \(T\) time units, we call \(T\) the
\emph{completion time} and \(K/T\) the \emph{All-Reduce rate}.
We index time units by \(t=0,1,\ldots,T-1\). A symbol sent in time unit \(t\)
is received at the end of that time unit.

The classical ring All-Reduce algorithm consists of a \emph{reduction} phase and a \emph{multicast} phase.
First consider \(K=1\).
The reduction phase takes \(N-1\) time units: in time unit \(t\), for
\(0\le t\le N-2\), node \(t\) sends the partial sum \(\sum_{i=0}^{t}x_{i,0}\) to node \(t+1\), which adds \(x_{t+1,0}\) to the received value.
At the end of time unit \(N-2\), node \(N-1\) holds \(\sum_{i=0}^{N-1}x_{i,0}\).
In the multicast phase, node \(N-1\) sends the sum to node \(0\), which forwards it around the ring; this likewise takes \(N-1\) time units.
Thus one sum symbol is computed and delivered to every node in \(T=2(N-1)\) time units, yielding a rate of \(\frac{1}{2(N-1)}\).

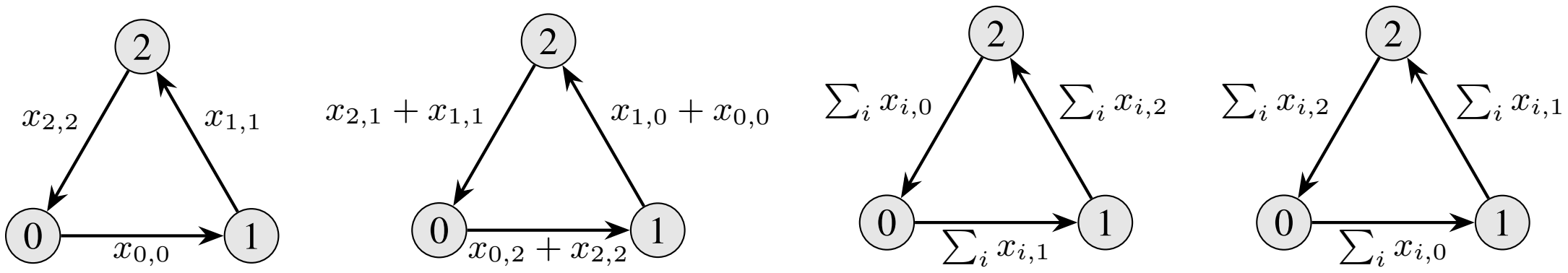
\begin{figure}[t]
    \centering
    \subfloat[\footnotesize Time unit $0$]{
        \begin{tikzpicture}
            \node[ringcir] (p1) at (-1,0) {0};
            \node[ringcir] (p2) at (1,0) {1};
            \node[ringcir] (p3) at (0,{sqrt(3)}) {2};
            \draw[ringedge] (p1) -- (p2) node[ringlab] {$x_{0,0}$};
            \draw[ringedge] (p2) -- (p3) node[ringlab] {$x_{1,1}$};
            \draw[ringedge] (p3) -- (p1) node[ringlab] {$x_{2,2}$};
        \end{tikzpicture}
    }
    \vspace{0.2cm}
    \subfloat[\footnotesize Time unit $1$]{
        \begin{tikzpicture}
            \node[ringcir] (p1) at (-1,0) {0};
            \node[ringcir] (p2) at (1,0) {1};
            \node[ringcir] (p3) at (0,{sqrt(3)}) {2};
            \draw[ringedge] (p1) -- (p2) node[ringlab] {$x_{0,2}+x_{2,2}$};
            \draw[ringedge] (p2) -- (p3) node[ringlab] {$x_{1,0}+x_{0,0}$};
            \draw[ringedge] (p3) -- (p1) node[ringlab] {$x_{2,1}+x_{1,1}$};
        \end{tikzpicture}
    }
    \vspace{0.2cm}
    \subfloat[\footnotesize Time unit $2$]{
        \begin{tikzpicture}
            \node[ringcir] (p1) at (-1,0) {0};
            \node[ringcir] (p2) at (1,0) {1};
            \node[ringcir] (p3) at (0,{sqrt(3)}) {2};
            \draw[ringedge] (p1) -- (p2) node[ringlab] {$\sum_{i}x_{i,1}$};
            \draw[ringedge] (p2) -- (p3) node[ringlab] {$\sum_{i}x_{i,2}$};
            \draw[ringedge] (p3) -- (p1) node[ringlab] {$\sum_{i}x_{i,0}$};
        \end{tikzpicture}
    }
    \vspace{0.2cm}
    \subfloat[\footnotesize Time unit $3$]{
        \begin{tikzpicture}
            \node[ringcir] (p1) at (-1,0) {0};
            \node[ringcir] (p2) at (1,0) {1};
            \node[ringcir] (p3) at (0,{sqrt(3)}) {2};
            \draw[ringedge] (p1) -- (p2) node[ringlab] {$\sum_{i}x_{i,0}$};
            \draw[ringedge] (p2) -- (p3) node[ringlab] {$\sum_{i}x_{i,1}$};
            \draw[ringedge] (p3) -- (p1) node[ringlab] {$\sum_{i}x_{i,2}$};
        \end{tikzpicture}
    }
    \caption{Classical ring All-Reduce on a $3$-node ring with $K=3$: three sums in $T=4$ time units.}
    \label{fig:ring_all_reduce_example}
\end{figure}

To improve throughput, the classical ring All-Reduce algorithm pipelines \(K=N\) coordinate-wise summation tasks, initiated at different nodes.
The two phases are commonly called \emph{Reduce-Scatter} and \emph{All-Gather}.
Together they compute and disseminate \(N\) sums in \(T=2(N-1)\) time units, yielding a rate of \(\frac{N}{2(N-1)}\).
Fig.~\ref{fig:ring_all_reduce_example} illustrates the schedule on a three-node ring.

\begin{figure}[t]
    \centering
    \subfloat[\footnotesize Time unit $0$]{
        \begin{tikzpicture}
            \node[ringcir] (p1) at (-1,0) {0};
            \node[ringcir] (p2) at (1,0) {1};
            \node[ringcir] (p3) at (0,{sqrt(3)}) {2};
            \draw[ringedge] (p1) -- (p2) node[ringlab] {$x_{0,1}$};
            \draw[ringedge] (p2) -- (p3) node[ringlab] {$x_{1,0}+x_{1,1}$};
            \draw[ringedge] (p3) -- (p1) node[ringlab] {$x_{2,0}$};
        \end{tikzpicture}
    }
    \vspace{0.35cm}
    \subfloat[\footnotesize Time unit $1$]{
        \begin{tikzpicture}
            \node[ringcir] (p1) at (-1,0) {0};
            \node[ringcir] (p2) at (1,0) {1};
            \node[ringcir] (p3) at (0,{sqrt(3)}) {2};
            \draw[ringedge] (p1) -- (p2) node[ringlab] {$x_{0,0}+x_{2,0}$};
            \draw[ringedge] (p2) -- (p3) node[ringlab] {$x_{0,1}+x_{1,1}$};
            \draw[ringedge] (p3) -- (p1) node[ringlab] {$\sum_{i=1}^{2}(x_{i,0}+x_{i,1})$};
        \end{tikzpicture}
    }
    \vspace{0.35cm}
    \subfloat[\footnotesize Time unit $2$]{
        \begin{tikzpicture}
            \node[ringcir] (p1) at (-1,0) {0};
            \node[ringcir] (p2) at (1,0) {1};
            \node[ringcir] (p3) at (0,{sqrt(3)}) {2};
            \draw[ringedge] (p1) -- (p2) node[ringlab, below=5pt] {$\sum_{i=0}^{2}(x_{i,0}+x_{i,1})$};
            \draw[ringedge] (p2) -- (p3) node[ringlab] {$\sum_{i=0}^{2}x_{i,0}$};
            \draw[ringedge] (p3) -- (p1) node[ringlab] {$\sum_{i=0}^{2}x_{i,1}$};
        \end{tikzpicture}
    }
    \caption{A linear network code for a three-node ring with \(K=2\).
    At time unit \(2\), node \(2\) receives \(\sum_{i=0}^{2}x_{i,0}\) and sends its previously computed sum
    \(\sum_{i=0}^{2}x_{i,1}\), while node \(0\) sends
    \(\sum_{i=0}^{2}(x_{i,0}+x_{i,1})\).
    Nodes \(0\) and \(1\) recover the missing coordinate sums by subtraction, so every node obtains both sums in \(T=3\) time units.}
    \label{fig:23_example}
\end{figure}
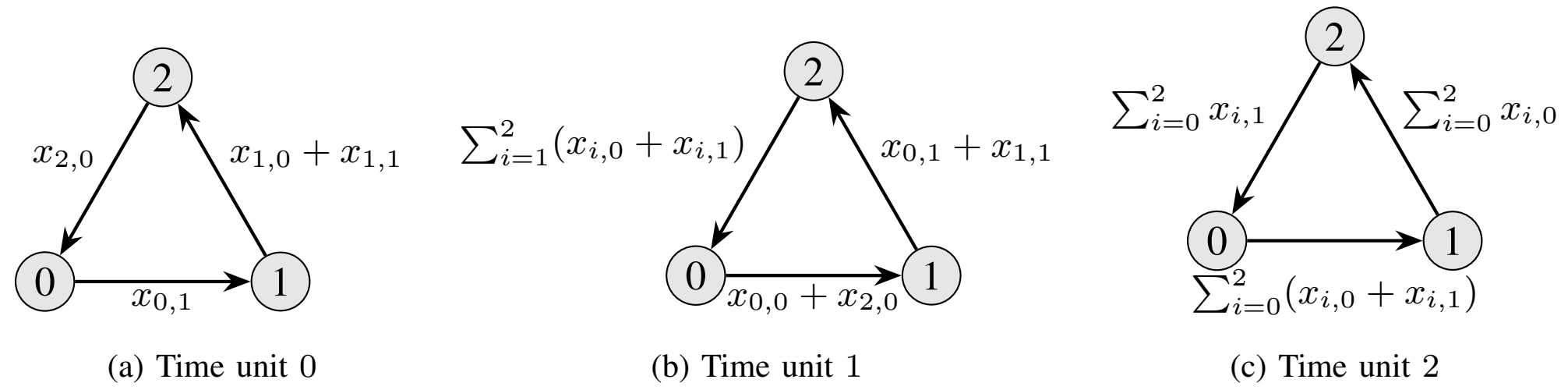

The classical algorithm combines only symbols with the same coordinate index.
A natural question is whether joint coding across coordinates can improve the
completion time for some \((K,N)\).
Consider the three-node ring in Fig.~\ref{fig:3-node-ring} with \(K=2\).
Fig.~\ref{fig:23_example} shows a linear network code that finishes in
\(T=3\) time units.
As shown later, no reduce--multicast code can achieve this completion time.

In the remainder of this paper, we study linear network codes for sum All-Reduce
on rings that allow joint coding across coordinates and thus extend beyond
reduce--multicast.

\section{Linear Network Coding for All-Reduce over a Ring Network}
\label{sec:linear_network_coding}

In this section, we develop a linear network coding framework for sum All-Reduce on the \(N\)-node directed ring \(G=(\mathcal V,\mathcal E)\). Because \(G\) contains a directed cycle, convolutional network codes~\cite[Chapter~20]{yeung2010information} are a natural option, but they use time-invariant coefficients. We instead form the time-expanded network \(G_T\) and apply a (possibly time-varying) linear network code on this directed acyclic graph, which makes causality explicit and simplifies the algebraic analysis.

\subsection{Time-Expanded Network}
\begin{definition}[Time-Expanded Network]
Let \(G=(\mathcal V,\mathcal E)\) be a directed network, and let \(T\geq 1\) be an integer. The \emph{time-expanded network} of \(G\) over \(T\) time units is the directed graph $G_T=(\mathcal V_T,\mathcal E_T)$,
defined as follows. Its vertex set is
\[
\mathcal V_T=\bigcup_{t=0}^{T}\mathcal V^{(t)},
\qquad
\mathcal V^{(t)}=\{i^{(t)}:i\in\mathcal V\},
\]
where \(i^{(t)}\) represents node \(i\) at time \(t\). The edge set
\(\mathcal E_T\) contains the following two types of edges:
\begin{itemize}
    \item \textbf{Transmission edges:}
    For every \((i,j)\in\mathcal E\) and every
    \(t\in\{0,1,\ldots,T-1\}\), the edge $(i^{(t)},j^{(t+1)})$
    is included in \(\mathcal E_T\) and has the same capacity as
    \((i,j)\) in \(G\).

    \item \textbf{Storage edges:}
    For every \(i\in\mathcal V\) and every
    \(t\in\{0,1,\ldots,T-1\}\), the edge $(i^{(t)},i^{(t+1)})$
    is included in \(\mathcal E_T\) with infinite capacity.
\end{itemize}
\end{definition}

The network $G_T$ models the propagation of data through $G$ over $T$ time units. A transmission edge represents the delivery of a
symbol over a physical edge during one time unit, while a storage edge represents a node retaining all information available to it for use at
a later time. Since each edge in \(G_T\) connects a vertex at time
\(t\) to one at time \(t+1\), \(G_T\) is directed and acyclic.

The \(K\)-symbol sum All-Reduce problem over \(G\) can be casted as the following multi-sum problem over \(G_T\):
\begin{itemize}
    \item The vertices in \(\mathcal V^{(0)}\) are the source vertices,
    with each \(i^{(0)}\) holding the vector
    \(\mathbf x_i\in\mathbb F^K\).

    \item The vertices in \(\mathcal V^{(T)}\) are the destination
    vertices, and every \(i^{(T)}\) must recover
    \(\sum_{j\in\mathcal V}\mathbf x_j\).
\end{itemize}

\begin{figure}[t]
    \centering
    \begin{tikzpicture}[
        scale=0.65,
        font=\footnotesize,
        cir/.style={
            circle,
            draw,
            fill=gray!30,
            minimum size=0.4cm,
            inner sep=0pt
        },
        edgelabel/.style={
            midway,
            inner sep=1.5pt,
            font=\scriptsize
        },
        lastlabel/.style={
            pos=0.60,
            below,
            sloped,
            fill=white,
            fill opacity=0.92,
            text opacity=1,
            inner sep=1.5pt,
            font=\scriptsize
        },
        transmission/.style={
            ->,
            >={Stealth[scale=1.0]}
        },
        storage/.style={
            ->,
            >={Stealth[scale=1.0]},
            dashed,
            gray
        }
    ]

        \begin{scope}[xshift=0cm]
            \node[cir] (p10) at (0,2) {$0^{(0)}$};
            \node[cir] (p20) at (0,0) {$1^{(0)}$};
            \node[cir] (p30) at (0,-2) {$2^{(0)}$};
        \end{scope}

        \begin{scope}[xshift=3cm]
            \node[cir] (p11) at (0,2) {$0^{(1)}$};
            \node[cir] (p21) at (0,0) {$1^{(1)}$};
            \node[cir] (p31) at (0,-2) {$2^{(1)}$};
        \end{scope}

        \begin{scope}[xshift=6cm]
            \node[cir] (p12) at (0,2) {$0^{(2)}$};
            \node[cir] (p22) at (0,0) {$1^{(2)}$};
            \node[cir] (p32) at (0,-2) {$2^{(2)}$};
        \end{scope}

        \begin{scope}[xshift=9cm]
            \node[cir] (p13) at (0,2) {$0^{(3)}$};
            \node[cir] (p23) at (0,0) {$1^{(3)}$};
            \node[cir] (p33) at (0,-2) {$2^{(3)}$};
        \end{scope}

        \begin{scope}[xshift=12cm]
            \node[cir] (p14) at (0,2) {$0^{(4)}$};
            \node[cir] (p24) at (0,0) {$1^{(4)}$};
            \node[cir] (p34) at (0,-2) {$2^{(4)}$};
        \end{scope}

\node[font=\small, above=0.2cm of p10] {\(t=0\)};
\node[font=\small, above=0.2cm of p11] {\(t=1\)};
\node[font=\small, above=0.2cm of p12] {\(t=2\)};
\node[font=\small, above=0.2cm of p13] {\(t=3\)};
\node[font=\small, above=0.2cm of p14] {\(t=4\)};
        \draw[storage] (p10) -- (p11);
        \draw[storage] (p11) -- (p12);
        \draw[storage] (p12) -- (p13);
        \draw[storage] (p13) -- (p14);

        \draw[storage] (p20) -- (p21);
        \draw[storage] (p21) -- (p22);
        \draw[storage] (p22) -- (p23);
        \draw[storage] (p23) -- (p24);

        \draw[storage] (p30) -- (p31);
        \draw[storage] (p31) -- (p32);
        \draw[storage] (p32) -- (p33);
        \draw[storage] (p33) -- (p34);

        \draw[transmission]
            (p10) -- (p21)
            node[edgelabel, above, sloped] {$x_{0,0}$};

        \draw[transmission]
            (p20) -- (p31)
            node[lastlabel] {$x_{1,1}$};

        \draw[transmission]
            (p30) -- (p11)
            node[edgelabel, above, sloped] {$x_{2,2}$};

        \draw[transmission]
            (p11) -- (p22)
            node[edgelabel, above, sloped]
            {$x_{0,2}+x_{2,2}$};

        \draw[transmission]
            (p21) -- (p32)
            node[lastlabel]
            {$x_{1,0}+x_{0,0}$};

        \draw[transmission]
            (p31) -- (p12)
            node[edgelabel, above, sloped]
            {$x_{2,1}+x_{1,1}$};

        \draw[transmission]
            (p12) -- (p23)
            node[edgelabel, above, sloped]
            {$\sum_{i}x_{i,1}$};

        \draw[transmission]
            (p22) -- (p33)
            node[lastlabel]
            {$\sum_{i}x_{i,2}$};

        \draw[transmission]
            (p32) -- (p13)
            node[edgelabel, above, sloped]
            {$\sum_{i}x_{i,0}$};

        \draw[transmission]
            (p13) -- (p24)
            node[edgelabel,above,sloped]
            {$\sum_{i}x_{i,0}$};

        \draw[transmission]
            (p23) -- (p34)
            node[lastlabel]
            {$\sum_{i}x_{i,1}$};

        \draw[transmission]
            (p33) -- (p14)
            node[edgelabel,above,sloped]
            {$\sum_{i}x_{i,2}$};

    \end{tikzpicture}

    \caption{Time-expanded network for a three-node ring network with
    \(K=3\) and \(T=4\). Solid and dashed edges represent transmission
    and storage edges, respectively.}
    \label{fig:example_time_expanded_network}
\end{figure}
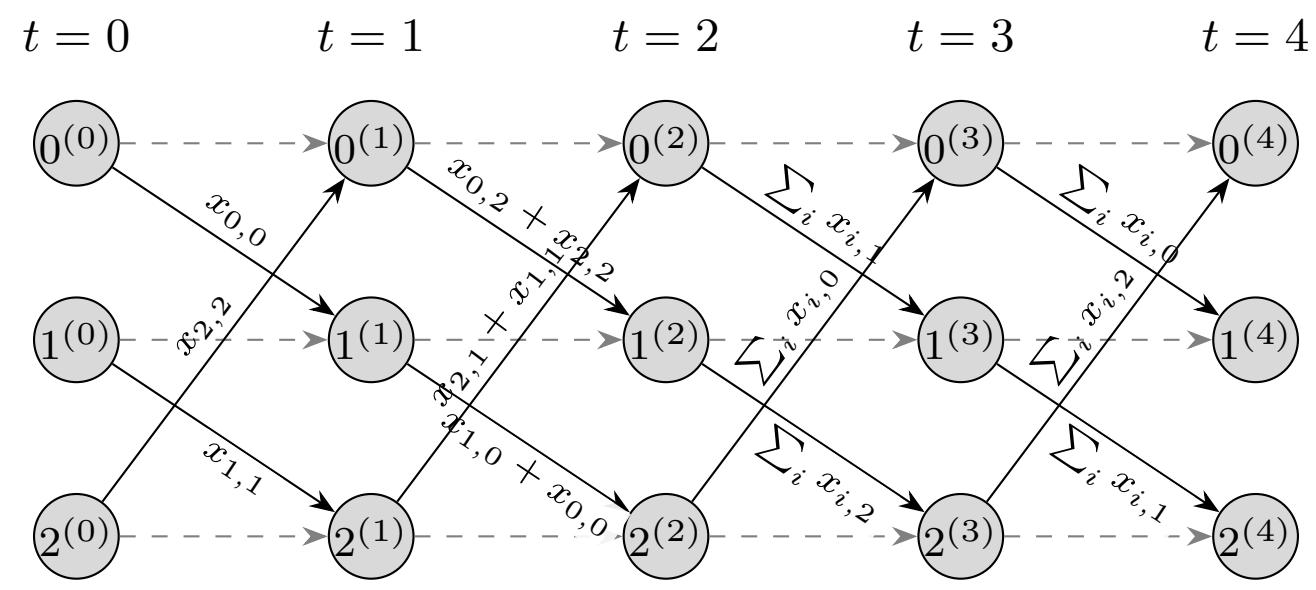

Therefore, any All-Reduce algorithm operating on \(G\) for \(T\) time units
induces a solution to the corresponding multi-sum problem on \(G_T\).
Conversely, a solution on \(G_T\) specifies a causal \(T\)-time-unit
All-Reduce algorithm on \(G\). For example, the ring All-Reduce
algorithm illustrated in Fig.~\ref{fig:ring_all_reduce_example} can be
represented on the time-expanded network in
Fig.~\ref{fig:example_time_expanded_network} as a linear network code
whose coding coefficients belong to \(\{0,1\}\). In the remainder of
this section, we study linear network codes for the multi-sum problem on
\(G_T\) with arbitrary coding coefficients in \(\mathbb F\).

\subsection{Linear Network Coding Formulation}
A linear network code consists of source encoding, linear recoding of received symbols, and decoding of
the desired sum at each destination. Throughout, message vectors are
represented as column vectors.

Let \(y_{i\rightarrow}^{(t)}\) denote the symbol transmitted by node \(i\) in time unit \(t\), and let \(y_{i\leftarrow}^{(t)}\) denote the symbol received at the beginning of time unit \(t\), equivalently
at the end of time unit \(t-1\), both through  transmission edges. Writing \(i'=(i-1)\bmod N\) for the predecessor of \(i\),
\begin{equation}
    y_{i\leftarrow}^{(t)}
    =
    y_{i'\rightarrow}^{(t-1)},
    \quad
    t\in\{1,\ldots,T\}.
    \label{equ:y}
\end{equation}

For each \(t\in\{1,\ldots,T\}\), the received-symbol vector at node \(i\) up to time index \(t\) is
\begin{equation}
    \mathbf{s}_i^{(t)}
    =
    \begin{bmatrix}
        y_{i\leftarrow}^{(1)} &
        \cdots &
        y_{i\leftarrow}^{(t)}
    \end{bmatrix}^{\mathsf T}
    \in\mathbb F^{t}.\label{equ:s}
\end{equation}
We set \(\mathbf{s}_i^{(0)}=\emptyset\) (equivalently, the empty vector).
In time unit \(t\in\{1,\ldots,T-1\}\), node \(i\) has access to \(\mathbf{x}_i\) and \(\mathbf{s}_i^{(t)}\), and transmits
\begin{equation}\label{eq:shuffle}
    y_{i\rightarrow}^{(t)}
    =
    \mathbf{m}_i^{(t)}\mathbf{x}_i
    +
    \boldsymbol{\lambda}_i^{(t)}\mathbf{s}_i^{(t)},
\end{equation}
where \(\mathbf{m}_i^{(t)}\in\mathbb F^{1\times K}\) and \(\boldsymbol{\lambda}_i^{(t)}\in\mathbb F^{1\times t}\) are the source-encoding and recoding coefficient vectors.
In time unit \(t=0\), before any symbols have been received, node \(i\) transmits
\begin{equation}
    y_{i\rightarrow}^{(0)}
    =
    \mathbf{m}_i^{(0)}\mathbf{x}_i,
    \quad
    \mathbf{m}_i^{(0)}\in\mathbb F^{1\times K}.
    \label{eq:initial_encoding}
\end{equation}

At the destination layer \(i^{(T)}\), node \(i\) forms the estimate
\begin{equation}
    \hat{\mathbf z}_i
    =
    \mathbf R_i\mathbf{s}_i^{(T)}
    +
    \mathbf D_i\mathbf{x}_i,
    \label{eq:decoding}
\end{equation}
from \(\mathbf{s}_i^{(T)}\) and its local message \(\mathbf{x}_i\), where \(\mathbf R_i\in\mathbb F^{K\times T}\) and \(\mathbf D_i\in\mathbb F^{K\times K}\) are decoding matrices.

Reduce--multicast codes form a special class in which each transmitted symbol depends on only one message coordinate, and each coordinate of the sum is recovered from a single received symbol together with at most one local source symbol. Section~\ref{sec:rm-codes} develops the broader class of \emph{coordinate-wise} linear network codes, which includes reduce--multicast as a special case.

\begin{definition}[Feasibility]
A \((K,T)\) linear network code for $K$-symbol sum All-Reduce on the $N$-node directed ring, specified by
\[
\left\{
\mathbf m_i^{(t)}:0\leq t\leq T-1;\,
\boldsymbol{\lambda}_i^{(t)}:1\leq t\leq T-1;\,
\mathbf R_i,\mathbf D_i
\right\}_{i\in\mathcal V}
\]
as in \eqref{eq:shuffle}, \eqref{eq:initial_encoding}, and \eqref{eq:decoding}, is \emph{feasible} if for every collection of source vectors \(\{\mathbf x_j\in\mathbb F^K:j\in\mathcal V\}\) and every destination \(i\in\mathcal V\),
\[
\hat{\mathbf z}_i
=
\sum_{j\in\mathcal V}\mathbf x_j.
\]
\end{definition}

The rate of a \((K,T)\) linear network code is \(K/T\).
A rate \(R\) is \emph{achievable} if for every \(\epsilon>0\) there exists a feasible \((K,T)\) linear network code with \(K/T\ge R-\epsilon\).

\subsection{A Necessary and Sufficient Condition for Feasibility}

For each node \(i\in\mathcal V\), collect the source-encoding coefficients into
\[
\mathbf M_i
=
\begin{bmatrix}
\mathbf m_i^{(0)}\\
\mathbf m_i^{(1)}\\
\vdots\\
\mathbf m_i^{(T-1)}
\end{bmatrix}
\in\mathbb F^{T\times K}.
\]
Define the recoding matrix
\(\boldsymbol{\Lambda}_i\in\mathbb F^{T\times T}\) by padding each
recoding vector with trailing zeros:
\[
[\boldsymbol{\Lambda}_i]_{t,:}
=
\begin{bmatrix}
\boldsymbol{\lambda}_i^{(t)} &
\mathbf 0_{1\times(T-t)}
\end{bmatrix},
\qquad 0\leq t\leq T-1,
\]
where $[\boldsymbol{\Lambda}_i]_{t,:}$ denotes its \(t\)-th row, rows are indexed from \(0\), and
\(\boldsymbol{\lambda}_i^{(0)}\) is the empty vector. In particular,
the first row of $\boldsymbol{\Lambda}_i$ is zero. $\boldsymbol{\Lambda}_i$ is strictly lower triangular,
reflecting causality: the recoding term in slot \(t\) uses only
symbols received at the beginnings of time units \(1,\ldots,t\),
which were transmitted in time units \(0,\ldots,t-1\).

For each \(i\in\mathcal V\) and each nonnegative integer
\(\ell\in\mathbb Z_{\geq 0}\) define
\begin{equation} \label{eq:A-definition}
\mathbf A_{i,\ell}
=
\boldsymbol{\Lambda}_i
\boldsymbol{\Lambda}_{i-1}
\cdots
\boldsymbol{\Lambda}_{i-\ell+1},
\end{equation}
where all node indices are interpreted modulo \(N\), and
\(\mathbf A_{i,0}=\mathbf I_T\). In particular,
\[
\mathbf A_{i,N}
=
\boldsymbol{\Lambda}_i
\boldsymbol{\Lambda}_{i-1}
\cdots
\boldsymbol{\Lambda}_{i-N+1}
\]
is the product corresponding to one full traversal of the \(N\)-node ring. 

\begin{theorem} \label{theorem:necsuf}
A $(K,T)$ linear network code for sum All-Reduce over the $N$-node directed
ring is feasible if and only if its coefficient matrices satisfy:
\begin{align}
\mathbf R'_{i+1}\mathbf A_{i,h}\mathbf M_{i-h}
&=
\mathbf I_K,\quad
\forall\, i\in\mathcal V,\quad
h\in\{0,\ldots,N-2\},
\label{equ:identity}
\\
\mathbf R'_{i+1}\mathbf A_{i,N-1}\mathbf M_{i+1}
+\mathbf D_{i+1}
&=
\mathbf I_K,\quad 
\forall\, i\in\mathcal V,
\label{equ:txi}
\end{align}
where
\[
\mathbf R'_{i+1}
=
\mathbf R_{i+1}
\left(
\mathbf I_T-\mathbf A_{i,N}
\right)^{-1}.
\]
The inverse is well defined as $\mathbf A_{i,N}$ is strictly lower triangular. All node indices above are
understood modulo $N$.
\end{theorem}

\begin{remark}
\label{remark:essential-feasibility}
The main feasibility requirements are the identities in
\eqref{equ:identity} for the nonlocal sources. 
They involve
only the matrices
$\{\mathbf M_i,\boldsymbol{\Lambda}_i,\mathbf R_i\}_{i\in\mathcal V}$.
Once matrices satisfying \eqref{equ:identity} are found, the
local-decoding matrices are uniquely determined by
\begin{equation}
\label{equ:D}
    \mathbf D_{i+1}
=
\mathbf I_K
-
\mathbf R'_{i+1}
\mathbf A_{i,N-1}
\mathbf M_{i+1},
\qquad i\in\mathcal V.
\end{equation}
Thus \eqref{equ:txi} imposes no additional constraint when the matrices
$\mathbf D_i$ are unrestricted.
\end{remark}

\begin{remark}
For each \(i\in\mathcal V\), \eqref{equ:identity} requires the same
matrix \(\mathbf R'_{i+1}\) to be a common left inverse of the \(N-1\)
transfer matrices $\mathbf A_{i,h}\mathbf M_{i-h}, h\in\{0,\ldots,N-2\}$.
Since $\mathbf R'_{i+1}
=
\mathbf R_{i+1}
(\mathbf I_T-\mathbf A_{i,N})^{-1}$
and \(\mathbf I_T-\mathbf A_{i,N}\) is invertible, specifying
\(\mathbf R'_{i+1}\) is equivalent to specifying
\(\mathbf R_{i+1}\). Thus, \eqref{equ:identity} jointly
constrains the decoder and all $N-1$ transfer matrices.

This condition is stronger than the full-column rank requirement in
single-source multicast, where the decoder can be chosen as a left
inverse of a single 
source-to-destination transfer matrix. Here, one decoder must invert
\(N-1\) transfer matrices simultaneously. Therefore, feasibility cannot
be reduced to the separate   nonsingularity conditions
$\det(\mathbf R'_{i+1}\mathbf A_{i,h}\mathbf M_{i-h})\neq 0$ for all $i$ and $h$.
These conditions ensure only that each product is invertible, not that
all products equal \(\mathbf I_K\), as required by
\eqref{equ:identity}.

 This distinction also explains why the standard random-coding argument
for multicast does not apply directly. Since \(\mathbf A_{i,N}\) is
strictly lower triangular, it is nilpotent, and
\begin{equation}
\label{equ:nilpotent}
    (\mathbf I_T-\mathbf A_{i,N})^{-1}
=
\sum_{t=0}^{T-1}\mathbf A_{i,N}^{t}.
\end{equation}
Thus, every entry of
\(\mathbf R'_{i+1}\mathbf A_{i,h}\mathbf M_{i-h}=\mathbf R_{i+1}
(\mathbf I_T-\mathbf A_{i,N})^{-1}\mathbf A_{i,h}\mathbf M_{i-h}\) is a polynomial in
the coding and decoding coefficients. Let \(p\) denote any diagonal
entry of this product, viewed as such a polynomial. Feasibility
requires \(p=1\). When all coefficients are zero, \(p=0\). Hence \(p-1\) is a nonzero
polynomial. 
If all coding and decoding coefficients are chosen independently and
uniformly from a finite field \(\mathbb F\), the Schwartz--Zippel
lemma gives
\[
\Pr[p=1]=\Pr[p-1=0]\leq \frac{\zeta}{|\mathbb F|},
\]
where \(\zeta\) bounds the degree of \(p-1\). Thus, for fixed \(K\) and \(T\), independently choosing all encoding
and decoding coefficients uniformly at random yields a feasible
All-Reduce code with probability tending to zero as the field size \(|\mathbb F|\)
grows. This differs from standard random linear multicast coding, in
which random encoding typically produces a full-rank transfer matrix
over a sufficiently large field, after which a suitable decoder is
chosen.
\end{remark}

The proof of Theorem~\ref{theorem:necsuf} is provided in Appendix~\ref{appendix:necsuf}. 
Substituting \eqref{equ:nilpotent} into its feasibility conditions yields an equivalent time-unrolled form, which is derived in Appendix~\ref{section:TUFC}.  In the following sections,
we use the feasibility condition in Theorem~\ref{theorem:necsuf} to derive an upper bound on the rate of a feasible code and construct feasible codes.

\section{Bounds on Completion Time and Rate of Feasible Codes}
\label{sec:rate-upper-bounds}
In this section, we derive lower bounds on the completion time \(T\)
for a fixed message length \(K\), and hence upper bounds on the rate
\(K/T\). We first obtain a bound from maximum flows in the
time-expanded network. We then derive a stronger bound using the
feasibility conditions in Theorem~\ref{theorem:necsuf}.

\subsection{Maximum-Flow Bound}
Any feasible \((K,T)\) linear network code for the multi-sum problem on \(G_T\) must support
multicast from each source to all destinations. Indeed, if all source
vectors except \(\mathbf{x}_i\) are set to zero, every destination must
recover \(\mathbf{x}_i\).

Let \(\operatorname{MaxFlow}(i^{(0)},j^{(T)})\) denote the maximum
flow value from \(i^{(0)}\) to \(j^{(T)}\) in \(G_T\).
By the max-flow min-cut
theorem, this equals the capacity of a minimum cut separating the two
vertices. For \(i\neq j\), since transmission edges have unit capacity, this value also equals the maximum
number of paths from \(i^{(0)}\) to \(j^{(T)}\) that share no
transmission edge. These paths may share storage edges.
Recovering the \(K\) symbols in \(\mathbf{x}_i\) requires
\[
\operatorname{MaxFlow}(i^{(0)},j^{(T)})\geq K
\]
for every \(i,j\in\mathcal V\). Evaluating these maximum flows yields
the following necessary condition.

\begin{proposition}\label{prop:necessary}
Let \(N\geq 3\) and \(K\geq 1\). A feasible \((K,T)\) linear network
code for sum All-Reduce on the \(N\)-node directed ring must satisfy
\[
T\geq K+N-2.
\]
\end{proposition}

\begin{proof}
All node indices are understood modulo $N$. Fix a source \(i^{(0)}\) and the destination \(j^{(T)}\), where
\(j=i-1\). If all source vectors other than \(\mathbf{x}_i\)
are set to zero, feasibility requires \(j^{(T)}\) to recover the \(K\)
symbols in \(\mathbf{x}_i\). Thus, $\operatorname{MaxFlow}(i^{(0)},j^{(T)})\geq K$.

Every path from \(i^{(0)}\) to \(j^{(T)}\) must traverse at least
\(N-1\) transmission edges. If \(T<N-1\), no such path exists,
contradicting feasibility. Otherwise, the first transmission on any
such path must occur in a time unit \(t\leq T-N+1\). Hence, every path
uses one of the following \(T-N+2\) unit-capacity edges:
\[
\left\{
\bigl(i^{(t)},(i+1)^{(t+1)}\bigr)
:
0\leq t\leq T-N+1
\right\}.
\]
Removing these edges disconnects \(i^{(0)}\) from \(j^{(T)}\).
By the max-flow min-cut theorem,
\[
K
\leq \operatorname{MaxFlow}(i^{(0)},j^{(T)})
\leq T-N+2.
\]
Rearranging gives \(T\geq K+N-2\).
\end{proof}

Proposition~\ref{prop:necessary} implies that every feasible $(K,T)$ linear network code satisfies \(\frac{K}{T} \leq \frac{K}{K+N-2} < 1\). When $K=N$, this gives $\frac{K}{T} \leq\frac{N}{2(N-1)}$. The classical ring All-Reduce algorithm attains this bound and is thus optimal for $K=N$. For fixed \(N\), however,
\(\frac{K}{K+N-2}\) approaches \(1\) as \(K\) grows. Thus, this bound alone
does not give a nontrivial upper bound on the supremum rate when
\(N\geq 3\).

\subsection{Linear-Coding Bound}

\begin{theorem}
    \label{theorem:necessary}
    Consider an \(N\)-node directed ring with \(N\geq 3\). If a feasible
    \((K,T)\) linear network code exists for sum All-Reduce over this ring, then
    \begin{equation}
        \label{eq:combined-converse}
        T \geq
        \max\left\{
        K+N-2,\,
        \left\lceil \frac{2(N-1)}{N}K \right\rceil
        \right\}.
        \end{equation}
\end{theorem}
For \(K\leq N\), the bound in
Theorem~\ref{theorem:necessary} reduces to
\(T\geq K+N-2\), which coincides with the maximum-flow bound in Proposition~\ref{prop:necessary}. For \(K>N\), it reduces to
\(
T\geq
\left\lceil\frac{2(N-1)}{N}K\right\rceil\),
which is strictly stronger than the maximum-flow bound.
The theorem implies that every feasible \((K,T)\) linear network code satisfies
\[
\frac{K}{T}
\leq
\min\left\{
\frac{K}{K+N-2},\,
\frac{K}{\left\lceil\frac{2(N-1)}{N}K\right\rceil}
\right\}
\leq
\frac{N}{2(N-1)}.
\]
The classical ring All-Reduce algorithm uses \(K=N\) and \(T=2(N-1)\), achieving rate \(\frac{N}{2(N-1)}\). Hence this is the supremum linear-coding rate on a directed ring; general linear network coding cannot improve it.
For a fixed \(K\), whether a code attains the completion-time lower bound, and whether coordinate-wise coding suffices, are addressed in Sections~\ref{sec:reduce-multicast-sufficiency} and~\ref{sec:construct-network-coding}.

Proposition~\ref{prop:necessary} establishes the first term in
\eqref{eq:combined-converse}. To prove
Theorem~\ref{theorem:necessary}, it therefore remains to show that
every feasible code satisfies
\begin{equation}
\label{eq:rate-converse}
T \geq \frac{2(N-1)}{N}K.
\end{equation}
Since \(T\) is an integer, this yields the second term in
\eqref{eq:combined-converse}. We prove \eqref{eq:rate-converse} below.

The two bounds arise from different arguments. Proposition~\ref{prop:necessary} uses the capacity of a cut in \(G_T\) for a single source--destination pair. By contrast, the proof of \eqref{eq:rate-converse} uses the feasibility identities in Theorem~\ref{theorem:necsuf}. These identities require one decoder to be a left inverse of \(N-1\) transfer matrices simultaneously. By bounding the dimension of the decoder’s kernel, we obtain a stronger constraint on \(T\).

According to Theorem~\ref{theorem:necsuf}, feasibility requires for every $i\in\mathcal V$ and every
$h\in\{0,1,\ldots,N-2\}$,
\begin{equation}
\label{eq:feas_general}
\mathbf R'_{i+1}
\mathbf A_{i,h}
\mathbf M_{i-h}
=
\mathbf I_K.
\end{equation}
In particular, taking $h=0$ and using $\mathbf A_{i,0}=\mathbf I_T$ gives
\begin{equation*}
\mathbf R'_{i+1}\mathbf M_i
=
\mathbf I_K.
\end{equation*}
Thus
\[
\operatorname{rank}(\mathbf M_i)=K, \quad 
\operatorname{rank}(\mathbf R'_{i+1})=K.
\]

Define the rank deficit
\begin{equation*}
d=T-K\ge 0
\end{equation*}
and the null space
\begin{equation*}
\mathcal N_{i+1}
=
\ker\mathbf R'_{i+1}
\subseteq\mathbb F^T.
\end{equation*}
Since $\mathbf R'_{i+1}$ is a $K\times T$ matrix of rank $K$, the
rank--nullity theorem gives
\begin{equation*}
\dim\mathcal N_{i+1}
=
T-\operatorname{rank}(\mathbf R'_{i+1})
=
T-K
=
d.
\end{equation*}

We prove \eqref{eq:rate-converse} separately for $N=3$ and $N=4$, and then
give a uniform argument for all $N\ge 5$.

\input{N3N4}

\input{N5andbeyond}

\section{Coordinate-Wise Codes and Completion-Time Achievability}
\label{sec:reduce-multicast-sufficiency}
In this section, we study \emph{coordinate-wise linear network codes}
and determine when they attain the completion-time lower bound in
Theorem~\ref{theorem:necessary}. This class includes reduce--multicast
codes, such as the classical ring All-Reduce construction. We introduce
two variants, interleaved reduce--multicast codes and
reduce--partial-multicast codes, that attain the bound for specified
parameter pairs $(K,T)$.

\subsection{Coordinate-wise Linear Network Codes}
\label{sec:rm-codes}
A \emph{coordinate-wise linear network code} processes each symbol
coordinate independently: no encoding, recoding, or decoding operation
combines symbols associated with different coordinates. In particular,
each transmitted symbol is associated with at most one coordinate, and
the coefficient vectors and matrices satisfy the following properties:
\begin{itemize}
    \item Each source-encoding vector \(\mathbf m_i^{(t)}\) has at most
    one nonzero entry. Thus, at most one coordinate of
    \(\mathbf x_i\) is incorporated into the transmitted symbol
    \(y_{i\rightarrow}^{(t)}\).

    \item Each recoding vector \(\boldsymbol{\lambda}_i^{(t)}\) has at
    most one nonzero entry. Hence, at most one previously received
    symbol is incorporated into \(y_{i\rightarrow}^{(t)}\).
    If both \(\mathbf m_i^{(t)}\) and \(\boldsymbol{\lambda}_i^{(t)}\)
    are nonzero, they select symbols associated with the same coordinate.

    \item For each output coordinate \(k\), row \(k\) of \(\mathbf D_i\)
    has at most one nonzero entry, while row \(k\) of \(\mathbf R_i\)
    may have multiple nonzero entries. Every symbol selected by these
    rows must be associated with coordinate \(k\).
\end{itemize}

A \emph{reduce--multicast code} processes each symbol coordinate
independently: it first combines the nodes' local symbols to form the
coordinate sum, then forwards that sum to the remaining nodes. It is
a coordinate-wise linear network code with the following additional
restrictions:
\begin{itemize}
    \item All entries of \(\mathbf m_i^{(t)}\),
    \(\boldsymbol{\lambda}_i^{(t)}\), \(\mathbf R_i\), and
    \(\mathbf D_i\) belong to \(\{0,1\}\subseteq\mathbb F\).
    \item Each row of \(\mathbf R_i\) has at most one nonzero entry.
    Thus, each coordinate sum is recovered from at most one received
    symbol, possibly combined with the local source symbol for that
    coordinate.
\end{itemize}
The classical ring All-Reduce algorithm yields a reduce--multicast code,
as described in Section~\ref{sec:ring_all}.

\subsection{Achievability for $K\geq N$}

Fix \(N\geq 3\). For \(K\geq N\), \(\frac{2(N-1)}{N}K\geq K+N-2\), so
the completion-time lower bound in Theorem~\ref{theorem:necessary} becomes
\[
T\geq T^*
=
\left\lceil\frac{2(N-1)}{N}K\right\rceil.
\]
We construct reduce--multicast codes that attain \(T^*\).

We first introduce the \emph{direct-sum} construction. Given feasible
\((K_1,T_1)\) and \((K_2,T_2)\) codes, run the first code on the first
\(K_1\) symbol coordinates during time units \(0,\ldots,T_1-1\),
then run the second on the remaining \(K_2\) coordinates during time units
\(T_1,\ldots,T_1+T_2-1\). The result is a feasible
\((K_1+K_2,T_1+T_2)\) code. If both codes are
reduce--multicast codes, so is their direct sum. The same construction
applies to any finite number of codes.

Write \(K=qN+N+r\) with integers \(q\geq 0\) and \(0\leq r<N\).
Then
\[
T^*
=
2(N-1)q
+
2(N+r-1)
-
\left\lfloor\frac{2r}{N}\right\rfloor.
\]
The direct sum of \(q\) copies of the classical
\((N,2(N-1))\) ring code from
Section~\ref{sec:ring_all} handles the first \(qN\) coordinates
(this component is omitted when \(q=0\)). It therefore remains to
construct a reduce--multicast code with parameters
\[
\left(
N+r,\,
2(N+r-1)-\left\lfloor\frac{2r}{N}\right\rfloor
\right).
\]
Taking its direct sum with the \(q\) classical codes then gives a
\((K,T^*)\) reduce--multicast code. For \(r=0\), the remaining code is
itself the classical \((N,2(N-1))\) ring code. Hence only
\(1\leq r<N\) requires a new construction.

\begin{construction}[Interleaved reduce--multicast code]
\label{def:interleaved-reduce-multicast}
Fix \(N\geq 3\) and \(1\leq r<N\). The code processes \(N+r\) symbol
coordinates on the \(N\)-node directed ring. Each coordinate is
treated as a task requiring \(2(N-1)\) transmissions: \(N-1\) to form
its sum and \(N-1\) to distribute the sum.

\begin{enumerate}
    \item \textbf{Initial tasks.} In time unit \(0\), each node
    \(i=0,\ldots,N-1\) starts the task for coordinate \(i\) by
    transmitting \(x_{i,i}\).

    \item \textbf{Additional tasks.} In time unit \(1\), for each
    \(j=0,\ldots,r-1\), node
    \(\upsilon_j=\left\lfloor\frac{jN}{r}\right\rfloor\)
    starts the task for coordinate \(N+j\) by transmitting
    \(x_{\upsilon_j,N+j}\).
    It retains the task received from its predecessor in time unit \(0\).
    Every other node adds its local symbol to the partial sum received
    in time unit \(0\) and forwards the result.

    \item \textbf{Scheduling.} From time unit \(2\) onward, if a node
    has both a retained task and a newly received task, it transmits
    the one with more transmissions remaining and retains the other.
    A task that has already undergone \(h\) transmissions has
    \(2(N-1)-h\) transmissions remaining. Ties are resolved according
    to a fixed rule.

    \item \textbf{Processing.} During a task's reduction phase, a node
    adds its local symbol for that coordinate to the received partial
    sum before forwarding it. In particular, the recipient of the
    \((N-1)\)-st reduction transmission adds the final local
    contribution, so the task's value is the completed sum before
    distribution begins. During the distribution phase, nodes forward
    the completed sum unchanged. A task is removed from the schedule
    after its \(2(N-1)\)-st transmission.
\end{enumerate}
\end{construction}

Figure~\ref{fig:interleaved-n4} illustrates the schedules for \(N=4\)
and \(r=1,2,3\). To illustrate the benefit of interleaving, consider a
sequential direct-sum schedule. It first runs the classical ring code
for the initial \(N\) coordinates, completing after \(2(N-1)\) time
units. Only then does it run a separate reduce--multicast code for the
remaining \(r\) coordinates. If each of these \(r\) tasks takes
\(2(N-1)\) time units, the sequential schedule completes after
\(4(N-1)\) time units. Because \(r<N\), this second stage does not
use every ring edge in every time unit (see Fig.~\ref{fig:stacked-n4}).
The interleaved construction instead starts the additional \(r\) tasks
in time unit \(1\), while the initial tasks are still running. Their
transmissions share the ring, avoiding a separate second stage and
making better use of the available edge capacity.

\begin{figure}[t]
\centering
\subfloat[$r=1$]{\begin{minipage}[t]{0.25\columnwidth}
\centering
\footnotesize
\setlength{\tabcolsep}{0.42em}
\begin{tabular}{c|cccccccc}
$i\backslash t$ & 0 & 1 & 2 & 3 & 4 & 5 & 6 & 7 \\
\hline
0 & 0 & \extra{4} & 3 & 2
  & \fullsum{1} & \fullsum{0}
  & \fullsum{\extra{4}} & \fullsum{3} \\

1 & 1 & 0 & \extra{4} & 3
  & \fullsum{2} & \fullsum{1}
  & \fullsum{0} & \fullsum{\extra{4}} \\

2 & 2 & 1 & 0 & \extra{4}
  & \fullsum{3} & \fullsum{2} & \fullsum{1} & \\

3 & 3 & 2 & 1 & \fullsum{0}
  & \fullsum{\extra{4}} & \fullsum{3} & \fullsum{2} & \\
\end{tabular}
\end{minipage}} \quad 
\subfloat[$r=2$]{\begin{minipage}[t]{0.35\columnwidth}
\centering
\footnotesize
\setlength{\tabcolsep}{0.42em}
\begin{tabular}{c|ccccccccc}
$i\backslash t$ & 0 & 1 & 2 & 3 & 4 & 5 & 6 & 7 & 8 \\
\hline
0 & 0 & \extra{4} & 3 & 2 & \extra{5}
  & \fullsum{1} & \fullsum{0}
  & \fullsum{\extra{4}} & \fullsum{3} \\

1 & 1 & 0 & \extra{4} & 3 & \fullsum{2}
  & \fullsum{\extra{5}} & \fullsum{1}
  & \fullsum{0} & \fullsum{\extra{4}} \\

2 & 2 & \extra{5} & 1 & 0 & \extra{4}
  & \fullsum{3} & \fullsum{2}
  & \fullsum{\extra{5}} & \fullsum{1} \\

3 & 3 & 2 & \extra{5} & 1 & \fullsum{0}
  & \fullsum{\extra{4}} & \fullsum{3}
  & \fullsum{2} & \fullsum{\extra{5}} \\
\end{tabular}
\end{minipage}} \quad 
\subfloat[$r=3$]{\begin{minipage}[t]{0.3\columnwidth}
\centering
\footnotesize
\setlength{\tabcolsep}{0.35em}
\begin{tabular}{c|ccccccccccc}
$i\backslash t$ & 0 & 1 & 2 & 3 & 4 & 5 & 6 & 7 & 8 & 9 & 10 \\
\hline
0 & 0 & \extra{4} & 3 & 2 & \extra{6}
  & \fullsum{1}
  & \fullsum{\extra{5}}
  & \fullsum{0}
  & \fullsum{\extra{4}}
  & \fullsum{3}
  & \\

1 & 1 & \extra{5} & 0 & \extra{4} & 3
  & \fullsum{2}
  & \fullsum{\extra{6}}
  & \fullsum{1}
  & \fullsum{\extra{5}}
  & \fullsum{0}
  & \fullsum{\extra{4}} \\

2 & 2 & \extra{6} & 1 & \extra{5} & 0
  & \extra{4}
  & \fullsum{3}
  & \fullsum{2}
  & \fullsum{\extra{6}}
  & \fullsum{1}
  & \fullsum{\extra{5}} \\

3 & 3 & 2 & \extra{6} & 1 & \extra{5}
  & \fullsum{0}
  & \fullsum{\extra{4}}
  & \fullsum{3}
  & \fullsum{2}
  & \fullsum{\extra{6}}
  & \\
\end{tabular}
\end{minipage}}
\caption{Interleaved reduce--multicast schedules for \(N=4\).
The entry in row \(i\) and column \(t\) is the coordinate transmitted
by node \(i\) in time unit \(t\); a blank entry means that the node is
idle. Nodes \(0,1,2,3\) inject coordinates \(0,1,2,3\), respectively,
in time unit \(0\). Blue boldface identifies transmissions of the additional
coordinates, which are injected in time unit \(1\). When two coordinates
have equal priority, the node transmits the one it retained. An unboxed entry denotes
a partial sum, while a boxed entry denotes the full coordinate sum.}
\label{fig:interleaved-n4}
\end{figure}

\begin{figure}[t]
\centering
\subfloat[$r=1$]{\begin{minipage}[t]{0.4\columnwidth}
\centering
\footnotesize
\setlength{\tabcolsep}{0.32em}
\begin{tabular}{c|cccccccccccc}
$i\backslash t$
  & 0 & 1 & 2 & 3 & 4 & 5 & 6 & 7 & 8 & 9 & 10 & 11 \\
\hline
0 & 0 & 3 & 2
  & \fullsum{1} & \fullsum{0} & \fullsum{3}
  & \extra{4} & & & & \fullsum{\extra{4}} & \\

1 & 1 & 0 & 3
  & \fullsum{2} & \fullsum{1} & \fullsum{0}
  & & \extra{4} & & & & \fullsum{\extra{4}} \\

2 & 2 & 1 & 0
  & \fullsum{3} & \fullsum{2} & \fullsum{1}
  & & & \extra{4} & & & \\

3 & 3 & 2 & 1
  & \fullsum{0} & \fullsum{3} & \fullsum{2}
  & & & & \fullsum{\extra{4}} & & \\
\end{tabular}
\end{minipage}}\quad 
\subfloat[$r=2$]{\begin{minipage}[t]{0.4\columnwidth}
\centering
\footnotesize
\setlength{\tabcolsep}{0.32em}
\begin{tabular}{c|cccccccccccc}
$i\backslash t$ & 0 & 1 & 2 & 3 & 4 & 5
               & 6 & 7 & 8 & 9 & 10 & 11 \\
\hline
0 & 0 & 3 & 2 & \fullsum{1} & \fullsum{0} & \fullsum{3}
  & \extra{4} & & \extra{5} &
  & \fullsum{\extra{4}} & \\

1 & 1 & 0 & 3 & \fullsum{2} & \fullsum{1} & \fullsum{0}
  & & \extra{4} & & \extra{5}
  & & \fullsum{\extra{4}} \\

2 & 2 & 1 & 0 & \fullsum{3} & \fullsum{2} & \fullsum{1}
  & \extra{5} & & \extra{4} &
  & \fullsum{\extra{5}} & \\

3 & 3 & 2 & 1 & \fullsum{0} & \fullsum{3} & \fullsum{2}
  & & \extra{5} & & \extra{4}
  & & \fullsum{\extra{5}} \\
\end{tabular}
\end{minipage}}
\caption{Sequential reduce--multicast schedule for \(N=4\). Coordinates
\(0,1,2,3\) start in time unit \(0\) and finish after six time units. The
bold coordinates start in time unit \(6\). Blank cells indicate idle nodes. An unboxed entry denotes a partial sum, while a boxed entry denotes the full coordinate sum.}
\label{fig:stacked-n4}
\end{figure}

\begin{theorem}\label{theorem:interleaved}
Let \(N\geq 3\) and \(1\leq r<N\). On the \(N\)-node directed ring,
the interleaved reduce--multicast construction yields a feasible code
for \(K=N+r\) symbols with completion time
\[
T=2(N+r-1)-\left\lfloor\frac{2r}{N}\right\rfloor.
\]
\end{theorem}

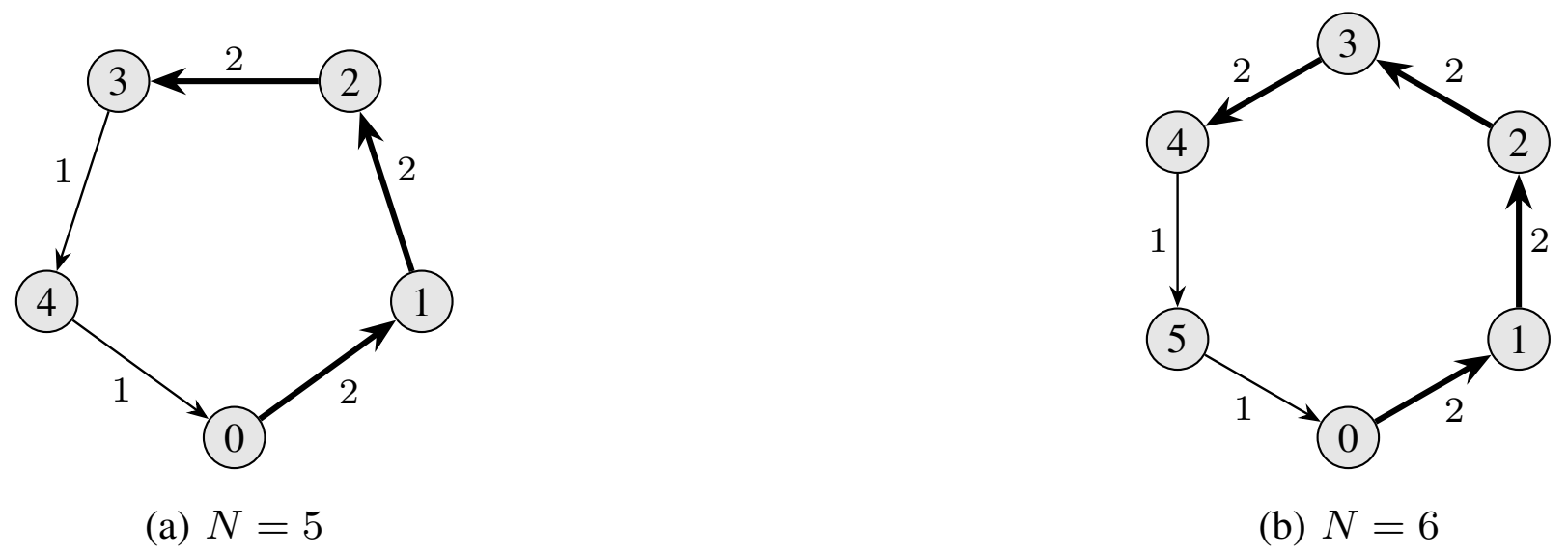
\begin{figure}[t]
    \centering
    \subfloat[$N=5$]{\begin{minipage}[t]{0.42\columnwidth}
    \centering
    \begin{tikzpicture}[
      cir/.style={circle, draw, fill=gray!20, minimum size=0.42cm, inner sep=1pt, font=\footnotesize},
      twice/.style={->, >={Stealth[scale=0.85]}, line width=1.15pt},
      once/.style={->, >={Stealth[scale=0.85]}, line width=0.45pt},
      elab/.style={midway, auto=right, font=\scriptsize, inner sep=1.5pt}
    ]
      \node[cir] (p0) at (270:1.35) {0};
      \node[cir] (p1) at (342:1.35) {1};
      \node[cir] (p2) at (54:1.35) {2};
      \node[cir] (p3) at (126:1.35) {3};
      \node[cir] (p4) at (198:1.35) {4};
      \draw[twice] (p0) -- (p1) node[elab] {$2$};
      \draw[twice] (p1) -- (p2) node[elab] {$2$};
      \draw[twice] (p2) -- (p3) node[elab] {$2$};
      \draw[once]  (p3) -- (p4) node[elab] {$1$};
      \draw[once]  (p4) -- (p0) node[elab] {$1$};
    \end{tikzpicture}
    \end{minipage}} \qquad
    \subfloat[$N=6$]{\begin{minipage}[t]{0.42\columnwidth}
    \centering
    \begin{tikzpicture}[
      cir/.style={circle, draw, fill=gray!20, minimum size=0.42cm, inner sep=1pt, font=\footnotesize},
      twice/.style={->, >={Stealth[scale=0.85]}, line width=1.15pt},
      once/.style={->, >={Stealth[scale=0.85]}, line width=0.45pt},
      elab/.style={midway, auto=right, font=\scriptsize, inner sep=1.5pt}
    ]
      \node[cir] (p0) at (270:1.35) {0};
      \node[cir] (p1) at (330:1.35) {1};
      \node[cir] (p2) at (30:1.35) {2};
      \node[cir] (p3) at (90:1.35) {3};
      \node[cir] (p4) at (150:1.35) {4};
      \node[cir] (p5) at (210:1.35) {5};
      \draw[twice] (p0) -- (p1) node[elab] {$2$};
      \draw[twice] (p1) -- (p2) node[elab] {$2$};
      \draw[twice] (p2) -- (p3) node[elab] {$2$};
      \draw[twice] (p3) -- (p4) node[elab] {$2$};
      \draw[once]  (p4) -- (p5) node[elab] {$1$};
      \draw[once]  (p5) -- (p0) node[elab] {$1$};
    \end{tikzpicture}
    \end{minipage}}
    \caption{Edge uses of the coordinate injected at node $0$. The label on each directed edge is the number of times that edge carries this coordinate. The $N-2$ edges that follow node $0$ are used twice, and the two edges entering node $0$ and its predecessor are used once.}
    \label{fig:one-symbol-edge-use}
    \end{figure}

\begin{proof}
All node indices are taken modulo \(N\).
In the construction, each coordinate follows successive edges of the directed ring,
pausing at a node only when another coordinate has priority. For a
coordinate originating at node \(i\), the first transmission carries
\(i\)'s local symbol. During the first \(N-1\) transmissions, each
successive recipient adds its local symbol, including the recipient
of the final reduction transmission. These recipients are precisely
the other \(N-1\) nodes, so the full sum is formed at node
\(i-1\). The next \(N-1\) transmissions forward this sum to all remaining nodes. Thus, \(2(N-1)\) transmissions
suffice to complete each coordinate task, regardless of any pauses.

If a task never completed, it would
eventually wait indefinitely at some node. The scheduling rule would then require
that node to transmit another unfinished task in every subsequent
time unit. This is impossible: there are finitely many tasks, and each
transmission reduces the total number of remaining transmissions by
one. Therefore, every task completes, and every node recovers every
coordinate sum. Hence, the code is feasible.

We next count the transmissions made by each node. A task started at node \(i\) requires two transmissions from every node
except the two nodes immediately preceding \(i\), namely \(i-1\)
and \(i-2\), which transmit it once each; see Fig.~\ref{fig:one-symbol-edge-use}.
Since the \(N\) tasks started in time unit \(0\) start at distinct
nodes, each node transmits \(N-2\) of these tasks twice
and the other two once, for a total of \(2(N-2)+2=2(N-1)\)
transmissions.

Let
\[
\Upsilon
=
\left\{
\left\lfloor\frac{jN}{r}\right\rfloor:0\leq j<r
\right\}
\]
be the set of nodes that start an additional task in time unit
\(1\). Each of these \(r\) tasks requires two transmissions
from node \(v\), unless it starts at \(v+1\) or \(v+2\), in which
case it requires only one. Thus, the total
number of transmissions made by node \(v\) is
\begin{equation}
\label{eq:total-transmissions}
L_v
=
2(N-1)+2r
-\mathbf 1[v+1\in\Upsilon]-\mathbf 1[v+2\in\Upsilon],
\end{equation}
where \(\mathbf 1[P]\) denotes the indicator of a statement \(P\): it equals \(1\) if \(P\) is true and \(0\) otherwise.

By Lemma~\ref{lemma:interleaved-contiguous} in Appendix~\ref{section:RMFC}, node \(v\) transmits in consecutive time units starting at time unit \(0\). Since it makes \(L_v\) transmissions, its last transmission occurs in time unit \(L_v-1\). The code therefore completes after \(\max_v L_v\) time units. Lemma~\ref{lemma:Lv-max} in Appendix~\ref{section:RMFC} gives \(\max_v L_v=2(N+r-1)-\lfloor 2r/N\rfloor\) as claimed.
\end{proof}

In summary, for every \(K\geq N\), direct sums of classical and
interleaved reduce--multicast codes attain the completion-time lower
bound.

\subsection{Achievability for $K<N$}
\label{sec:partial-multicast}

For \(N\geq 3\) and \(K<N\), \(\frac{2(N-1)}{N}K<K+N-2\), so the lower bound in Theorem~\ref{theorem:necessary} reduces to \(T\geq T^*=K+N-2\). The classical ring All-Reduce algorithm requires \(2(N-1)\) time units and hence does not attain \(T^*\).

To handle shorter message vectors, we extend the reduce--multicast construction by allowing \emph{partial multicast}. After a coordinate
sum is formed through \(N-1\) successive transmissions, it may be
forwarded to only the next \(n\) nodes, where \(0\leq n\leq N-1\).
The case \(n=N-1\) is full multicast. We investigate whether
reduce--multicast codes with this extension can attain \(T^*\).

\begin{construction}[Reduce--partial-multicast code]
\label{def:reduce-partial-multicast}
Let \(1\leq K<N\) and suppose that \(K\) divides \(N\). The
reduce--partial-multicast code is defined as follows:
\begin{enumerate}
    \item \textbf{Initialization.} In time unit \(0\), node \(i\) transmits
    its local symbol on coordinate \(i\bmod K\). Thus, each coordinate
    \(\kappa\) is started at the \(N/K\) nodes satisfying
    \(i\equiv\kappa\pmod K\).

    \item \textbf{Reduction.} In time units \(1,\ldots,N-2\), each node
    adds its local symbol for the received coordinate to the partial
    sum and forwards the result. The partial sums advance one node
    per time unit and remain spaced \(K\) nodes apart. At the end of
    time unit \(N-2\), each recipient adds its local symbol to complete
    the sum. Consequently, every node satisfying
    \(i\equiv\kappa-1\pmod K\) holds the full sum for coordinate
    \(\kappa\).

    \item \textbf{Partial multicast.} Each completed sum is forwarded
    unchanged to the next \(K-1\) nodes. This phase occupies time units
    \(N-1,\ldots,N+K-3\). For \(K=1\), this phase is omitted.
\end{enumerate}
\end{construction}

For each coordinate, the nodes holding its sum after reduction are
spaced \(K\) nodes apart. Forwarding each copy to the next \(K-1\)
nodes therefore delivers that sum to every node. Hence a feasible
reduce--partial-multicast code exists whenever \(K\) divides \(N\), and
it completes All-Reduce in \(T^*=N+K-2\) time units.
Figure~\ref{fig:k-divides-n} illustrates the schedules for \(N=6\)
with \(K=2\) and \(K=3\).

When \(K\) does not divide \(N\), however, no coordinate-wise linear
network code attains \(T^*\), as established by the following theorem.

\begin{figure}[t]
\centering
\subfloat[$K=2$]{\begin{minipage}[t]{0.46\columnwidth}
\centering
\footnotesize
\setlength{\tabcolsep}{0.45em}
\begin{tabular}{c|cccccc}
$i\backslash t$ & 0 & 1 & 2 & 3 & 4 & 5 \\
\hline
0 & 0 & 1 & 0 & 1 & 0 & \fullsum{1} \\
1 & 1 & 0 & 1 & 0 & 1 & \fullsum{0}\\
2 & 0 & 1 & 0 & 1 & 0 & \fullsum{1}\\
3 & 1 & 0 & 1 & 0 & 1 & \fullsum{0}\\
4 & 0 & 1 & 0 & 1 & 0 & \fullsum{1}\\
5 & 1 & 0 & 1 & 0 & 1 & \fullsum{0}\\
\end{tabular}
\end{minipage}}\quad
\subfloat[$K=3$]{\begin{minipage}[t]{0.48\columnwidth}
\centering
\footnotesize
\setlength{\tabcolsep}{0.42em}
\begin{tabular}{c|ccccccc}
$i\backslash t$ & 0 & 1 & 2 & 3 & 4 & 5 & 6 \\
\hline
0 & 0 & 2 & 1 & 0 & 2 & \fullsum{1} & \fullsum{0}\\
1 & 1 & 0 & 2 & 1 & 0 & \fullsum{2} & \fullsum{1} \\
2 & 2 & 1 & 0 & 2 & 1 & \fullsum{0} & \fullsum{2} \\
3 & 0 & 2 & 1 & 0 & 2 & \fullsum{1} & \fullsum{0}\\
4 & 1 & 0 & 2 & 1 & 0 & \fullsum{2} & \fullsum{1}\\
5 & 2 & 1 & 0 & 2 & 1 & \fullsum{0} & \fullsum{2} \\
\end{tabular}
\end{minipage}}
\caption{Reduce--partial-multicast code for \(N=6\), with each full
coordinate sum forwarded to the next \(K-1\) nodes. The entry in row
\(i\) and column \(t\) identifies the coordinate transmitted by node
\(i\) in time unit \(t\). In time unit \(0\), node \(i\) transmits coordinate
\(i\bmod K\). Unboxed entries denote partial sums; boxed entries
denote full coordinate sums.}
\label{fig:k-divides-n}
\end{figure}

\begin{theorem}
\label{thm:k-less-n}
Consider an \(N\)-node directed ring with \(N\geq3\) and \(1\leq K<N\).
A feasible coordinate-wise linear network code with completion time
\(N+K-2\) exists if and only if \(K\) divides \(N\). When \(K\) divides
\(N\), this time is attained by a reduce--partial-multicast code;
otherwise, no coordinate-wise linear network code, including any
reduce--multicast code, attains it.
\end{theorem}

The proof is provided in Appendix~\ref{section:proof-k-less-n}.
Thus, for \(K<N\), coordinate-wise linear coding attains the minimum
completion time \(N+K-2\) exactly when \(K\) divides \(N\). Otherwise,
attaining this bound requires joint coding across symbol coordinates.

\section{Achievability through Joint Linear Coding}
\label{sec:construct-network-coding}

Fix \(N\geq3\), and let \(2\leq K<N\) be an integer that does not
divide \(N\).
By Theorem~\ref{thm:k-less-n}, no coordinate-wise linear code attains
the completion-time lower bound \(N+K-2\) for these parameters.
We construct a linear network code that attains this bound through joint
coding across symbol coordinates.

\begin{theorem}\label{thm:construction}
Consider an \(N\)-node directed ring with \(N\geq3\). For every integer
\(K\) satisfying \(2\leq K<N\) that does not divide \(N\), a feasible
\((K,N+K-2)\) linear network code for All-Reduce exists over every
field.
\end{theorem}

\subsection{Construction of a \((K,N+K-2)\) Linear Network Code}

Let \(K\) be an integer satisfying \(2\leq K<N\) that does not divide \(N\). Set
\[
\varrho=N\bmod K,
\qquad T=N+K-2.
\]
All node indices are taken modulo \(N\). Row and column indices
start from \(0\). For \(j\geq0\), let \(\mathbf e_j\) denote the standard basis vector with a \(1\) in position \(j\) (and zeros elsewhere), interpreted as a row or column vector as needed; its length is determined by context.

\begin{construction}[Joint linear network code]
\label{def:joint-linear-construction}
The code is specified by the following steps.

\begin{enumerate}
\item \textbf{Construct the initial encoding vectors.}
Let \(\mathbf W\in\mathbb F^{\varrho\times K}\) collect the initial encoding
vectors of nodes \(0,\ldots,\varrho-1\).
Initialize its entries by
\[
\mathbf W[i,j]
=
\begin{cases}
1, & i\equiv j\pmod \varrho,\\
0, & \text{otherwise},
\end{cases}
\qquad 0\leq i<\varrho,\quad 0\leq j<K.
\]
Let \(\xi=K\bmod \varrho\). If \(\xi>0\),  replace
the submatrix of $\mathbf W$ on rows \(\xi,\ldots,\varrho-1\) and columns
\(0,\ldots,\xi-1\) by \(\mathbf H(\varrho-\xi,\xi)\). For positive integers \(a,b\),
define
\begin{equation}
\label{equ:H}
\mathbf H(a,b)
=
\begin{cases}
\mathbf I_a, & a=b,\\[1mm]
\begin{bmatrix}
\mathbf H(a-b,b)\\
\mathbf I_b
\end{bmatrix}, & a>b,\\[6mm]
\begin{bmatrix}
\mathbf H(a,b-a) & \mathbf I_a
\end{bmatrix}, & a<b.
\end{cases}
\end{equation}
When \(\xi=0\), no replacement is needed.

For each node \(i\), choose
\[
\mathbf m_i^{(0)}
=
\begin{cases}
\mathbf W[i,:], & 0\leq i<\varrho,\\
\mathbf e_{i\bmod K}, & \varrho\leq i<N.
\end{cases}
\]
In time unit \(0\), node \(i\) transmits
\[
y_{i\rightarrow}^{(0)}=\mathbf m_i^{(0)}\mathbf x_i.
\]

\item \textbf{Reduce and forward.}
For each node \(i\), choose
\[
\mathbf M_i
=
\begin{bmatrix}
\mathbf m_i^{(0)}\\
\mathbf m_{i-1}^{(0)}\\
\vdots\\
\mathbf m_{i-N+1}^{(0)}\\
\mathbf 0_{(K-2)\times K}
\end{bmatrix}
\in\mathbb F^{T\times K},
\qquad
\boldsymbol{\Lambda}_i
=
\boldsymbol{\Delta}_T
=
\begin{bmatrix}
0 & 0 & \cdots & 0\\
1 & 0 & \cdots & 0\\
0 & 1 & \ddots & \vdots\\
\vdots & \ddots & \ddots & 0\\
0 & \cdots & 1 & 0
\end{bmatrix}
\in\mathbb F^{T\times T}.
\]
The zero block in
\(\mathbf M_i\) is empty when \(K=2\). $\boldsymbol{\Delta}_T$ is the one-step shift matrix.

The first \(N\) rows of \(\mathbf M_i\) specify the reduction phase:
\[
\mathbf m_i^{(t)}=\mathbf m_{i-t}^{(0)}=
\mathbf m_{i-1}^{(t-1)},
\qquad 0\leq t\leq N-1.
\]
In time units \(1,\ldots,N-1\), node \(i\) adds the corresponding linear
combination of its local symbols to the received partial sum and
transmits
\[
y_{i\rightarrow}^{(t)}
=
y_{i\leftarrow}^{(t)}
+
\mathbf m_{i-t}^{(0)}\mathbf x_i.
\]
The remaining \(K-2\) rows are zero, so in time units \(N,\ldots,T-1\),
node \(i\) forwards the completed linear combination unchanged:
\[
y_{i\rightarrow}^{(t)}=y_{i\leftarrow}^{(t)}.
\]
The shift matrix \(\boldsymbol{\Delta}_T\) selects the most recently
received symbol in every time unit \(t\geq1\), while its first row is zero
because no symbol has been received before time unit \(0\). Here,
\(y_{i\leftarrow}^{(t)}\) is received from the transmission in
time unit \(t-1\). If \(T=N\), the forwarding phase is empty.

\item \textbf{Decode.}
For each node \(i\), define
\begin{equation}
\label{equ:QQ}
    \mathbf Q_i
=
\begin{bmatrix}
\mathbf m_{i+1}^{(0)}\\
\mathbf m_i^{(0)}\\
\mathbf m_{i-1}^{(0)}\\
\vdots\\
\mathbf m_{i-K+2}^{(0)}
\end{bmatrix}
\in\mathbb F^{K\times K}.
\end{equation}
The construction of \(\mathbf W\) ensures that \(\mathbf Q_i\) is
invertible over every field, as proved in
Lemma~\ref{lem:Qi-invertibility}. Choose the decoding matrices as
\[
\mathbf R_i
=
\mathbf Q_i^{-1}
\begin{bmatrix}
\mathbf 0_{K\times(N-2)} & \mathbf I_K
\end{bmatrix}
\in\mathbb F^{K\times T},
\qquad
\mathbf D_i
=
\mathbf Q_i^{-1}
\begin{bmatrix}
\mathbf m_{i+1}^{(0)}\\
\mathbf 0_{(K-1)\times K}
\end{bmatrix}
\in\mathbb F^{K\times K}.
\]
Writing \(\mathbf s_i^{(T)}=[y_{i\leftarrow}^{(1)},\ldots,y_{i\leftarrow}^{(T)}]^{\mathsf T}\), the factor \([\mathbf 0_{K\times(N-2)}\ \mathbf I_K]\) in the form of \(\mathbf R_i\) selects the last \(K\) entries, namely the symbols received at times \(N-1,\ldots,T\). The matrix \(\mathbf D_i\) supplies the local contribution needed to complete the first of these symbols.

\end{enumerate}
\end{construction}

\begin{example}[Joint linear coding for \(N=7\) and \(K=4\)]
Here,
\[
\varrho=7\bmod4=3,\qquad
\xi=4\bmod3=1,\qquad
T=N+K-2=9.
\]
The corner block is
\[
\mathbf H(2,1)=
\begin{bmatrix}1\\1\end{bmatrix},
\]
giving the initial encoding matrix
\[
\mathbf W=
\begin{bmatrix}
1&0&0&1\\
1&1&0&0\\
1&0&1&0
\end{bmatrix}.
\]

\paragraph{Initial transmissions.}
In time unit \(0\), nodes \(0,\ldots,6\) transmit the following symbols:
\[
\begin{array}{c|ccccccc}
\text{Node} & 0 & 1 & 2 & 3 & 4 & 5 & 6\\ \hline
\text{Symbol}
& x_{0,0}+x_{0,3}
& x_{1,0}+x_{1,1}
& x_{2,0}+x_{2,2}
& x_{3,3}
& x_{4,0}
& x_{5,1}
& x_{6,2}
\end{array}
\]
Thus, the first three nodes jointly encode two coordinates, while
the remaining nodes transmit a single coordinate.

\paragraph{Reduction and forwarding.}
Each initial transmission follows successive edges of the ring.
In time units \(1,\ldots,6\), each recipient adds the same combination
of its local symbols and forwards the result. In time units \(7\) and
\(8\), the completed combination is forwarded unchanged.

For example, the path originating at node \(0\) carries the
combination of coordinates \(0\) and \(3\). Node \(t\) transmits
\[
\sum_{\ell=0}^{t}(x_{\ell,0}+x_{\ell,3}),
\qquad t=0,\ldots,6.
\]
Writing
\[
\mathbf z=\sum_{j=0}^{6}\mathbf x_j,
\]
node \(6\) therefore transmits \(z_0+z_3\) in time unit \(6\).
Nodes \(0\) and \(1\) forward this value in time units \(7\) and \(8\),
respectively.

\paragraph{Decoding.}
Each node uses its last four receptions, adding its local
contribution to the first reception. At node \(6\), these give
\[
\mathbf u_6=
\begin{bmatrix}
y_{6\leftarrow}^{(6)}+x_{6,0}+x_{6,3}\\
y_{6\leftarrow}^{(7)}\\
y_{6\leftarrow}^{(8)}\\
y_{6\leftarrow}^{(9)}
\end{bmatrix}
=
\begin{bmatrix}
z_0+z_3\\
z_2\\
z_1\\
z_0
\end{bmatrix}.
\]
Thus, node \(6\) obtains \(z_0,z_1,z_2\) directly and recovers
\[
z_3=u_{6,0}-u_{6,3}.
\]

At node \(0\), the corresponding combinations are
\[
\mathbf u_0=
\begin{bmatrix}
y_{0\leftarrow}^{(6)}+x_{0,0}+x_{0,1}\\
y_{0\leftarrow}^{(7)}\\
y_{0\leftarrow}^{(8)}\\
y_{0\leftarrow}^{(9)}
\end{bmatrix}
=
\begin{bmatrix}
z_0+z_1\\
z_0+z_3\\
z_2\\
z_1
\end{bmatrix}.
\]
Node \(0\) recovers
\[
z_1=u_{0,3},\qquad
z_2=u_{0,2},\qquad
z_0=u_{0,0}-u_{0,3},\qquad
z_3=u_{0,1}-z_0.
\]
These operations are valid over every field, including fields
of characteristic two.

The remaining nodes decode using their respective invertible
matrices \(\mathbf Q_i\). All four coordinate sums are available
at every node by the end of time unit \(8\), so the code completes in
nine time units.
\end{example}

\subsection{Feasibility of the Codes in Construction~\ref{def:joint-linear-construction}}

The feasibility proof relies on the following property of the
decoding matrices.

\begin{lemma}[Invertibility of the decoding matrices]
\label{lem:Qi-invertibility}
For every \(i\in\mathcal V\), the matrix \(\mathbf Q_i\) defined
in \eqref{equ:QQ} has determinant \(\pm1\). In particular,
\(\mathbf Q_i\) is invertible over every field.
\end{lemma}

The proof is given in Appendix~\ref{app:decoding-invertibility}.
It establishes the required determinant properties of the recursive
corner block and the initial encoding matrix \(\mathbf W\).

\begin{proof}[Proof of Theorem~\ref{thm:construction}]
Let
\[
\mathbf z=\sum_{j\in\mathcal V}\mathbf x_j,
\qquad T=N+K-2.
\]
We show that the constructed codes in Construction~\ref{def:joint-linear-construction} recover
\(\mathbf z\) at every node.

By induction on \(t\), the reduction rule gives
\[
y_{i\rightarrow}^{(t)}
=
\mathbf m_{i-t}^{(0)}
\sum_{\ell=0}^{t}\mathbf x_{i-t+\ell},
\qquad 0\leq t\leq N-1,
\]
where all node indices are taken modulo \(N\). Indeed, the identity
holds at \(t=0\). Each subsequent transmission adds the corresponding
linear combination of the transmitting node's local symbols to
the received partial sum. In particular,
\[
y_{i\rightarrow}^{(N-1)}
=
\mathbf m_{i-N+1}^{(0)}\mathbf z.
\]
The forwarding rule then yields
\[
y_{i\rightarrow}^{(t)}
=
\mathbf m_{i-t}^{(0)}\mathbf z,
\qquad N-1\leq t\leq T-1.
\]

Since
\(y_{i\leftarrow}^{(t)}=y_{i-1\rightarrow}^{(t-1)}\),
the reception at time \(N-1\) satisfies
\[
y_{i\leftarrow}^{(N-1)}
=
\mathbf m_{i+1}^{(0)}
\sum_{j\neq i}\mathbf x_j.
\]
Adding the local contribution gives
\[
y_{i\leftarrow}^{(N-1)}
+
\mathbf m_{i+1}^{(0)}\mathbf x_i
=
\mathbf m_{i+1}^{(0)}\mathbf z.
\]
The remaining receptions satisfy
\[
y_{i\leftarrow}^{(N+j)}
=
\mathbf m_{i-j}^{(0)}\mathbf z,
\qquad j=0,\ldots,K-2.
\]
Consequently,
\[
\begin{bmatrix}
y_{i\leftarrow}^{(N-1)}
    +\mathbf m_{i+1}^{(0)}\mathbf x_i\\
y_{i\leftarrow}^{(N)}\\
\vdots\\
y_{i\leftarrow}^{(T)}
\end{bmatrix}
=
\begin{bmatrix}
\mathbf m_{i+1}^{(0)}\\
\mathbf m_i^{(0)}\\
\vdots\\
\mathbf m_{i-K+2}^{(0)}
\end{bmatrix}\mathbf z
=
\mathbf Q_i\mathbf z.
\]

With the constructed decoding matrices
\[
\mathbf R_i
=
\mathbf Q_i^{-1}
\begin{bmatrix}
\mathbf 0_{K\times(N-2)} & \mathbf I_K
\end{bmatrix},
\qquad
\mathbf D_i
=
\mathbf Q_i^{-1}
\begin{bmatrix}
\mathbf m_{i+1}^{(0)}\\
\mathbf 0_{(K-1)\times K}
\end{bmatrix},
\]
node \(i\) therefore obtains
\[
\hat{\mathbf z}_i
=
\mathbf R_i\mathbf s_i^{(T)}
+
\mathbf D_i\mathbf x_i
=
\mathbf Q_i^{-1}\mathbf Q_i\mathbf z
=
\mathbf z.
\]
The inverse exists over every field by
Lemma~\ref{lem:Qi-invertibility}. The last required reception arrives
at the end of time unit \(T-1\), so every node recovers the sum within
\(T=N+K-2\) time units. Hence, the constructed code is feasible over
every field.
\end{proof}

\section{Concluding Remarks}
\label{sec:conclusion}

This paper has characterized the completion time and hence the rate of sum All-Reduce on a directed ring under linear network coding.
Several directions remain open beyond this setting. We do not claim that the ring is the optimal topology for All-Reduce in every scenario. The time-expanded viewpoint extends to general cyclic topologies, but a node may then have multiple incoming and outgoing transmission edges, so the encoding, recoding, and decoding maps become vector- and matrix-valued. An important question is whether some topologies admit a strict rate gain from linear coding over reduce--multicast.

\appendices
\input{Appendix}

\bibliographystyle{IEEEtran}
\bibliography{capacity}

\end{document}

%% file: N3N4.tex
\subsection{The Case $N=3$}
\label{subsec:upper-bound-N3}

For $N=3$, the feasibility identities
\eqref{eq:feas_general}, corresponding to $h=0$ and $h=1$, give
\begin{align}
\mathbf R'_{i+1}\mathbf M_i
&=
\mathbf I_K,
\label{eq:N3-feas-h0}
\\
\mathbf R'_{i+1}
\boldsymbol{\Lambda}_i\mathbf M_{i-1}
&=
\mathbf I_K,
\label{eq:N3-feas-h1}
\end{align}
for every $i\in\mathcal V$. All node indices in this subsection are
understood modulo $3$.

Define the one-step difference matrix
\begin{equation}
\label{eq:N3-def-Phi}
\boldsymbol{\Phi}_i
=
\mathbf M_i-\boldsymbol{\Lambda}_i\mathbf M_{i-1}
\in\mathbb F^{T\times K}.
\end{equation}
Subtracting \eqref{eq:N3-feas-h1} from \eqref{eq:N3-feas-h0} yields
\[
\mathbf R'_{i+1}\boldsymbol{\Phi}_i=\mathbf 0.
\]
Therefore, every column of $\boldsymbol{\Phi}_i$ lies in
$\mathcal N_{i+1}=\ker\mathbf R'_{i+1}$. Since
$\dim\mathcal N_{i+1}=d$, it follows that
\begin{equation*}
\operatorname{rank}(\boldsymbol{\Phi}_i)
\le d.
\end{equation*}

Rearranging \eqref{eq:N3-def-Phi}, we obtain the recurrence
\begin{equation*}
\mathbf M_i
=
\boldsymbol{\Lambda}_i\mathbf M_{i-1}
+
\boldsymbol{\Phi}_i.
\end{equation*}
Applying this recurrence successively at nodes $i-1$ and $i-2$ gives
\begin{align*}
\mathbf M_i
&=
\boldsymbol{\Lambda}_i
\left(
\boldsymbol{\Lambda}_{i-1}\mathbf M_{i-2}
+
\boldsymbol{\Phi}_{i-1}
\right)
+
\boldsymbol{\Phi}_i
=
\boldsymbol{\Lambda}_i\boldsymbol{\Lambda}_{i-1}
\left(
\boldsymbol{\Lambda}_{i-2}\mathbf M_{i-3}
+
\boldsymbol{\Phi}_{i-2}
\right)
+
\boldsymbol{\Lambda}_i\boldsymbol{\Phi}_{i-1}
+
\boldsymbol{\Phi}_i.
\end{align*}
Because $i-3\equiv i\pmod 3$, this becomes
\[
\mathbf M_i
=
\mathbf A_{i,3}\mathbf M_i
+
\mathbf A_{i,2}
\boldsymbol{\Phi}_{i-2}+
\boldsymbol{\Lambda}_i\boldsymbol{\Phi}_{i-1}+
\boldsymbol{\Phi}_i.
\]
Hence
\begin{equation}
\label{eq:N3-cycle-expansion}
\left(\mathbf I_T-\mathbf A_{i,3}\right)\mathbf M_i
=
\boldsymbol{\Phi}_i
+
\boldsymbol{\Lambda}_i\boldsymbol{\Phi}_{i-1}
+
\mathbf A_{i,2}
\boldsymbol{\Phi}_{i-2}.
\end{equation}

Since $\mathbf A_{i,3}$ is strictly lower triangular, 
$\mathbf I_T-\mathbf A_{i,3}$ is invertible. Left multiplication by an invertible matrix preserves rank. So,
\[
\operatorname{rank}
\left(
\left(\mathbf I_T-\mathbf A_{i,3}\right)\mathbf M_i
\right)
=
\operatorname{rank}(\mathbf M_i)
=
K.
\]
Applying rank subadditivity to \eqref{eq:N3-cycle-expansion}, and using
$\operatorname{rank}(\mathbf B\mathbf C)
\le\operatorname{rank}(\mathbf C)$, gives
\begin{align*}
K
\le
\operatorname{rank}(\boldsymbol{\Phi}_i)
+
\operatorname{rank}
\left(
\boldsymbol{\Lambda}_i\boldsymbol{\Phi}_{i-1}
\right)
+
\operatorname{rank}
\left(
\mathbf A_{i,2}
\boldsymbol{\Phi}_{i-2}
\right)
\le
\operatorname{rank}(\boldsymbol{\Phi}_i)
+
\operatorname{rank}(\boldsymbol{\Phi}_{i-1})
+
\operatorname{rank}(\boldsymbol{\Phi}_{i-2})
\le 3d.
\end{align*}
Consequently,
\[
d\ge\frac{K}{3}.
\]
Recalling that $d=T-K$, we obtain
\[
T\ge K+\frac{K}{3}=\frac{4K}{3}.
\]
This is exactly the bound $T\ge\frac{2(N-1)}{N}K$
for $N=3$.

\subsection{The Case $N=4$}
\label{subsec:upper-bound-N4}

For $N=4$, the feasibility identities
\eqref{eq:feas_general}, corresponding to $h=0,1,2$, are for every $i\in\mathcal V$,
\begin{align}
\mathbf R'_{i+1}\mathbf M_i
&=
\mathbf I_K,
\label{eq:N4-feas-h0}
\\
\mathbf R'_{i+1}
\boldsymbol{\Lambda}_i\mathbf M_{i-1}
&=
\mathbf I_K,
\label{eq:N4-feas-h1}
\\
\mathbf R'_{i+1}
\mathbf A_{i,2}\mathbf M_{i-2}
&=
\mathbf I_K.
\label{eq:N4-feas-h2}
\end{align}
Here all node indices are understood modulo
$4$. The argument below uses the identities for $h=0$ and $h=2$.

Define the two-step difference matrix
\begin{equation}
\label{eq:N4-def-Psi}
\boldsymbol{\Psi}_i
=
\mathbf M_i-\mathbf A_{i,2}\mathbf M_{i-2}.
\end{equation}
Subtracting \eqref{eq:N4-feas-h2} from \eqref{eq:N4-feas-h0} gives
\[
\mathbf R'_{i+1}\boldsymbol{\Psi}_i=\mathbf 0.
\]
Thus
\[
\operatorname{col}(\boldsymbol{\Psi}_i)
\subseteq
\mathcal N_{i+1}.
\]
Since $\dim\mathcal N_{i+1}=d$, we obtain
\begin{equation*}
\operatorname{rank}(\boldsymbol{\Psi}_i)
\le d.
\end{equation*}

Rearranging \eqref{eq:N4-def-Psi} gives
\begin{equation}
\label{eq:N4-M-recurrence}
\mathbf M_i
=
\mathbf A_{i,2}\mathbf M_{i-2}
+
\boldsymbol{\Psi}_i.
\end{equation}
Applying the same identity with $i$ replaced by $i-2$ yields
\[
\mathbf M_{i-2}
=
\mathbf A_{i-2,2}\mathbf M_{i-4}
+
\boldsymbol{\Psi}_{i-2} = \mathbf A_{i-2,2}\mathbf M_i
+
\boldsymbol{\Psi}_{i-2},
\]
where the second equality holds as $i-4\equiv i\pmod 4$.
Substitution into \eqref{eq:N4-M-recurrence} gives
\begin{align*}
\mathbf M_i
&=
\mathbf A_{i,2}
\left(
\mathbf A_{i-2,2}\mathbf M_i
+
\boldsymbol{\Psi}_{i-2}
\right)
+
\boldsymbol{\Psi}_i
=
\mathbf A_{i,4}\mathbf M_i
+
\mathbf A_{i,2}\boldsymbol{\Psi}_{i-2}
+
\boldsymbol{\Psi}_i,
\end{align*}
where
$\mathbf A_{i,2}\mathbf A_{i-2,2}
=
\mathbf A_{i,4}$.
Therefore,
\begin{equation}
\label{eq:N4-cycle-expansion}
\left(\mathbf I_T-\mathbf A_{i,4}\right)\mathbf M_i
=
\boldsymbol{\Psi}_i
+
\mathbf A_{i,2}\boldsymbol{\Psi}_{i-2}.
\end{equation}
Since $\mathbf A_{i,4}$ is strictly lower triangular,
$\mathbf I_T-\mathbf A_{i,4}$ is invertible. Hence
\[
\operatorname{rank}
\left(
\left(\mathbf I_T-\mathbf A_{i,4}\right)\mathbf M_i
\right)
=\operatorname{rank}(\mathbf M_i)=
K.
\]
Using \eqref{eq:N4-cycle-expansion}, rank subadditivity, and the fact that
left multiplication cannot increase rank, we obtain
\begin{align*}
K
\le
\operatorname{rank}(\boldsymbol{\Psi}_i)
+
\operatorname{rank}
\left(
\mathbf A_{i,2}\boldsymbol{\Psi}_{i-2}
\right)
\le
\operatorname{rank}(\boldsymbol{\Psi}_i)
+
\operatorname{rank}(\boldsymbol{\Psi}_{i-2})
\le 2d.
\end{align*}
 Therefore,
\[
d\ge\frac{K}{2}.
\]
Since $d=T-K$, it follows that
\[
T\ge K+\frac{K}{2}=\frac{3K}{2}.
\]
This is exactly the bound $T\ge\frac{2(N-1)}{N}K$
for $N=4$.

%% file: N5andbeyond.tex
\subsection{The General Case $N\ge 5$}
\label{subsec:upper-bound-general}

We now consider $N\ge 5$. All node indices in this
subsection are understood modulo $N$.

For each $i\in\mathcal V$, define the one-step difference matrix
\begin{equation}
\label{eq:Phi_def_general}
\boldsymbol{\Phi}_i
=
\mathbf M_i
-
\boldsymbol{\Lambda}_i\mathbf M_{i-1}
\in\mathbb F^{T\times K}.
\end{equation}
For every $h\in\{0,1,\ldots,N-3\}$, feasibility gives
\begin{align}
\mathbf R'_{i+1}
\mathbf A_{i,h}\mathbf M_{i-h}
&=
\mathbf I_K,
\label{eq:general-adjacent-feas-1}
\\
\mathbf R'_{i+1}
\mathbf A_{i,h+1}\mathbf M_{i-h-1}
&=
\mathbf I_K.
\label{eq:general-adjacent-feas-2}
\end{align}
Subtracting \eqref{eq:general-adjacent-feas-2} from
\eqref{eq:general-adjacent-feas-1}, and using
$\mathbf A_{i,h+1}
=
\mathbf A_{i,h}\boldsymbol{\Lambda}_{i-h},
$
we obtain
\begin{align*}
\mathbf 0
&=
\mathbf R'_{i+1}
\left(
\mathbf A_{i,h}\mathbf M_{i-h}
-
\mathbf A_{i,h+1}\mathbf M_{i-h-1}
\right)
\\
&=
\mathbf R'_{i+1}\mathbf A_{i,h}
\left(
\mathbf M_{i-h}
-
\boldsymbol{\Lambda}_{i-h}\mathbf M_{i-h-1}
\right)
\\
&=
\mathbf R'_{i+1}
\mathbf A_{i,h}\boldsymbol{\Phi}_{i-h}.
\end{align*}
Consequently,
\begin{equation}
\label{eq:contain_1_general}
\operatorname{col}
\left(
\mathbf A_{i,h}\boldsymbol{\Phi}_{i-h}
\right)
\subseteq
\mathcal N_{i+1},
\qquad
h=0,1,\ldots,N-3.
\end{equation}

For each $i\in\mathcal V$, define
\begin{equation}
\label{eq:mu_def_general}
\mu_i
=
\operatorname{rank}
\left[
\mathbf A_{i,2}\boldsymbol{\Phi}_{i-2}
\;\;
\mathbf A_{i,3}\boldsymbol{\Phi}_{i-3}
\;\;
\cdots
\;\;
\mathbf A_{i,N-3}\boldsymbol{\Phi}_{i-N+3}
\right].
\end{equation}
The matrix in \eqref{eq:mu_def_general} contains $N-4$ block columns,
each of size $T\times K$.

\begin{lemma}[Local-overlap inequality]
\label{lemma:local-overlap}
For every $i\in\mathcal V$,
\begin{equation}
\label{eq:local_overlap}
K\le 2d-\mu_i.
\end{equation}
\end{lemma}

\begin{proof}
Replacing $i$ by $i-2$ in \eqref{eq:contain_1_general} gives
\[
\operatorname{col}
\left(
\mathbf A_{i-2,h}\boldsymbol{\Phi}_{i-2-h}
\right)
\subseteq
\mathcal N_{i-1},
\qquad
h=0,1,\ldots,N-3.
\]
Applying the linear map $\mathbf A_{i,2}$ to both sides yields
\[
\operatorname{col}
\left(
\mathbf A_{i,2}\mathbf A_{i-2,h}
\boldsymbol{\Phi}_{i-2-h}
\right)
\subseteq
\mathbf A_{i,2}\mathcal N_{i-1}.
\]
Since $\mathbf A_{i,2}\mathbf A_{i-2,h}
=
\mathbf A_{i,h+2}$,
reindexing $h+2$ as $h$ gives
\begin{equation}
\label{eq:contain_2_general}
\operatorname{col}
\left(
\mathbf A_{i,h}\boldsymbol{\Phi}_{i-h}
\right)
\subseteq
\mathbf A_{i,2}\mathcal N_{i-1},
\qquad
h=2,3,\ldots,N-1.
\end{equation}

Rearranging \eqref{eq:Phi_def_general}, we have
\[
\mathbf M_i
=
\boldsymbol{\Lambda}_i\mathbf M_{i-1}
+
\boldsymbol{\Phi}_i.
\]
Iterating this identity once around the ring gives
\[
\mathbf M_i
=
\mathbf A_{i,N}\mathbf M_i
+
\sum_{h=0}^{N-1}
\mathbf A_{i,h}\boldsymbol{\Phi}_{i-h}.
\]
Therefore,
\begin{equation}
\label{eq:M_expand_general}
\left(
\mathbf I_T-\mathbf A_{i,N}
\right)\mathbf M_i
=
\sum_{h=0}^{N-1}
\mathbf A_{i,h}\boldsymbol{\Phi}_{i-h}.
\end{equation}

Since $\mathbf I_T-\mathbf A_{i,N}$ is invertible, we have
\begin{equation}
\label{eq:M_rank_general}
\operatorname{rank}
\left(
\left(\mathbf I_T-\mathbf A_{i,N}\right)\mathbf M_i
\right)
=\operatorname{rank}
\left(\mathbf M_i
\right)=
K.
\end{equation}

By \eqref{eq:contain_1_general}, the terms on the right-hand side of
\eqref{eq:M_expand_general} with $h=0,1,\ldots,N-3$
have their columns in $\mathcal N_{i+1}$. By
\eqref{eq:contain_2_general}, the terms with
$h=2,3,\ldots,N-1$
have their columns in $\mathbf A_{i,2}\mathcal N_{i-1}$. These two
ranges cover all $h\in\{0,1,\ldots,N-1\}$. Therefore,
\[
\operatorname{col}
\left(
\left(\mathbf I_T-\mathbf A_{i,N}\right)\mathbf M_i
\right)
\subseteq
\mathcal N_{i+1}+\mathbf A_{i,2}\mathcal N_{i-1}.
\]
Using \eqref{eq:M_rank_general}, we obtain
\begin{equation}
\label{eq:K_sumspace_general}
K
\le
\dim
\left(
\mathcal N_{i+1}+\mathbf A_{i,2}\mathcal N_{i-1}
\right).
\end{equation}

The terms with
$h=2,3,\ldots,N-3$
belong to both subspaces. Hence the column space of
\[
\left[
\mathbf A_{i,2}\boldsymbol{\Phi}_{i-2}
\;\;
A_{i,3}\boldsymbol{\Phi}_{i-3}
\;\;
\cdots
\;\;
\mathbf A_{i,N-3}\boldsymbol{\Phi}_{i-N+3}
\right]
\]
is contained in
$\mathcal N_{i+1}
\cap
\mathbf A_{i,2}\mathcal N_{i-1}$.
By the definition of $\mu_i$ in \eqref{eq:mu_def_general},
\begin{equation}
\label{eq:intersection_mu_general}
\dim
\left(
\mathcal N_{i+1}
\cap
\mathbf A_{i,2}\mathcal N_{i-1}
\right)
\ge
\mu_i.
\end{equation}

Since $\dim\left(\mathbf A_{i,2}\mathcal N_{i-1}\right)
\le
\dim\mathcal N_{i-1}
=
d$ and $\dim\mathcal N_{i+1}=d$, we have
\begin{align*}
\dim
\left(
\mathcal N_{i+1}+\mathbf A_{i,2}\mathcal N_{i-1}
\right)
&=
\dim\mathcal N_{i+1}
+
\dim\left(\mathbf A_{i,2}\mathcal N_{i-1}\right)
-
\dim
\left(
\mathcal N_{i+1}\cap\mathbf A_{i,2}\mathcal N_{i-1}
\right)
\\
&\le
d+d-\mu_i
\\
&=
2d-\mu_i.
\end{align*}
Combining this inequality with \eqref{eq:K_sumspace_general} proves
\[
K\le 2d-\mu_i. \qedhere
\]
\end{proof}

To complete the proof, it remains to derive a global lower
bound on the overlap parameters $\mu_i$. We first introduce a block operator
that simultaneously represents one-step propagation at every node.

Define the $NT\times NT$ block matrix
\begin{equation*}
\boldsymbol{\Theta}
=
\begin{bmatrix}
\mathbf 0 & \mathbf 0 & \cdots & \mathbf 0
    & \boldsymbol{\Lambda}_0
\\
\boldsymbol{\Lambda}_1 & \mathbf 0 & \cdots & \mathbf 0
    & \mathbf 0
\\
\mathbf 0 & \boldsymbol{\Lambda}_2 & \ddots & \mathbf 0
    & \mathbf 0
\\
\vdots & \ddots & \ddots & \ddots & \vdots
\\
\mathbf 0 & \mathbf 0 & \cdots
    & \boldsymbol{\Lambda}_{N-1} & \mathbf 0
\end{bmatrix},
\end{equation*}
where every block is of size $T\times T$.
It can be verified that 
\begin{equation}
\label{eq:Theta-N}
\boldsymbol{\Theta}^{N}
=
\operatorname{diag}
\left(
\mathbf A_{0,N},
\mathbf A_{1,N},
\ldots,
\mathbf A_{N-1,N}
\right).
\end{equation}
Each $\mathbf A_{i,N}$ is a product of strictly lower-triangular matrices
and is therefore itself strictly lower triangular. Since it is a
$T\times T$ matrix,
\(
\left(\mathbf A_{i,N}\right)^T
=
\mathbf 0_{T\times T}\).
It follows from \eqref{eq:Theta-N} that
\(
\boldsymbol{\Theta}^{NT}
=
\mathbf 0_{NT\times NT}\).
Therefore, $\boldsymbol{\Theta}$ is nilpotent and
$\mathbf I_{NT}-\boldsymbol{\Theta}$ is invertible.

For every integer $h\ge 0$, define the stacked transfer matrix
\begin{equation*}
\mathbf U_h
=
\begin{bmatrix}
\mathbf A_{0,h}\mathbf M_{-h}
\\
\mathbf A_{1,h}\mathbf M_{1-h}
\\
\vdots
\\
\mathbf A_{N-1,h}\mathbf M_{N-1-h}
\end{bmatrix}
\in\mathbb F^{NT\times K},
\end{equation*}
where all node indices are understood modulo $N$.
We index the \(T\times K\) row blocks of \(\mathbf U_h\) by
\(i\in\{0,\ldots,N-1\}\), so that
\[
[\mathbf U_h]_i=\mathbf A_{i,h}\mathbf M_{i-h}.
\]
 By the definitions of
$\boldsymbol{\Theta}$ and $\mathbf A_{i,h}$,
\begin{equation}
\label{eq:general-U-recurrence}
\mathbf U_{h+1}
=
\boldsymbol{\Theta}\mathbf U_h,
\qquad h\ge 0.
\end{equation}
Indeed, the $i$th $T\times K$ block of
$\boldsymbol{\Theta}\mathbf U_h$ is
\[
\boldsymbol{\Lambda}_i
\mathbf A_{i-1,h}\mathbf M_{i-h-1}
=
\mathbf A_{i,h+1}\mathbf M_{i-(h+1)}
=
[\mathbf U_{h+1}]_i,
\]
which proves \eqref{eq:general-U-recurrence}.

\begin{lemma}[Independence of the stacked transfer matrices]
\label{lemma:stacked-independence}
For every integer $q$ satisfying $2\le q\le N-3$,
the matrix
\[
\left[
\mathbf U_2
\;\;
\mathbf U_3
\;\;
\cdots
\;\;
\mathbf U_q
\right]\in\mathbb F^{NT\times (q-1)K}
\]
has full column rank $(q-1)K$. Equivalently, if
$\mathbf a_2,\ldots,\mathbf a_q\in\mathbb F^K$ satisfy
\begin{equation}
\label{eq:general-U-relation}
\sum_{h=2}^{q}\mathbf U_h\mathbf a_h
=
\mathbf 0_{NT\times 1},
\end{equation}
then $\mathbf a_2=\mathbf a_3=\cdots=\mathbf a_q
=
\mathbf 0_{K\times 1}$.
\end{lemma}

\begin{proof}
We proceed by induction on $q$.
For $q=2$, \eqref{eq:general-U-relation} becomes
\[
\mathbf U_2\mathbf a_2=\mathbf 0.
\]
Its $i$th $T$-dimensional block is
\[
\mathbf A_{i,2}\mathbf M_{i-2}\mathbf a_2=\mathbf 0.
\]
Left-multiplying by $\mathbf R'_{i+1}$ and using the feasibility identity $\mathbf R'_{i+1}\mathbf A_{i,2}\mathbf M_{i-2}
=
\mathbf I_K$
gives $\mathbf a_2=\mathbf 0$.

Now let $3\le q\le N-3$ and assume that the result holds with $q-1$ in
place of $q$. Suppose that
\[
\sum_{h=2}^{q}\mathbf U_h\mathbf a_h=\mathbf 0.
\]
The $i$th block of this relation is
\[
\sum_{h=2}^{q}
\mathbf A_{i,h}\mathbf M_{i-h}\mathbf a_h
=
\mathbf 0.
\]
Because $h\le q\le N-3$, the feasibility identities apply to every term.
Left-multiplying by $\mathbf R'_{i+1}$ therefore yields
\begin{equation}
\label{eq:general-U-coefficient-sum}
\sum_{h=2}^{q}\mathbf a_h=\mathbf 0.
\end{equation}

For $h=2,\ldots,q-1$, define the partial sums
\[
\mathbf b_h
=
\sum_{\ell=2}^{h}\mathbf a_\ell.
\]
Then $\mathbf a_2=\mathbf b_2$, $\mathbf a_h=\mathbf b_h-\mathbf b_{h-1}$ for $3\le h\le q-1$,
and, by \eqref{eq:general-U-coefficient-sum},
$\mathbf a_q=-\mathbf b_{q-1}$.
Consequently,
\begin{align}
\sum_{h=2}^{q}\mathbf U_h\mathbf a_h
=
\sum_{h=2}^{q-1}
\left(
\mathbf U_h-\mathbf U_{h+1}
\right)\mathbf b_h
=
\left(
\mathbf I_{NT}-\boldsymbol{\Theta}
\right)
\sum_{h=2}^{q-1}\mathbf U_h\mathbf b_h,
\label{eq:general-U-factorization}
\end{align}
where the second equality follows from
\eqref{eq:general-U-recurrence}.

The left-hand side of \eqref{eq:general-U-factorization} is zero. Since
$\mathbf I_{NT}-\boldsymbol{\Theta}$ is invertible,
\[
\sum_{h=2}^{q-1}\mathbf U_h\mathbf b_h
=
\mathbf 0.
\]
By the induction hypothesis,
\[
\mathbf b_2=\mathbf b_3=\cdots=\mathbf b_{q-1}
=
\mathbf 0.
\]
The relations between the $\mathbf a_h$ and the $\mathbf b_h$ then imply
\[
\mathbf a_2=\mathbf a_3=\cdots=\mathbf a_q
=
\mathbf 0.
\]
This completes the induction.
\end{proof}

\begin{lemma}[Global overlap inequality]
\label{lemma:global-overlap}
For $N\ge 5$,
\begin{equation}
\label{eq:general-global-mu}
\sum_{i=0}^{N-1}\mu_i
\ge
(N-4)K.
\end{equation}
\end{lemma}

\begin{proof}
Define the $NT\times (N-4)K$ matrix
\begin{equation}
\label{eq:general-Xi}
\boldsymbol{\Xi}
=
\left[
\mathbf U_2-\mathbf U_3
\;\;
\mathbf U_3-\mathbf U_4
\;\;
\cdots
\;\;
\mathbf U_{N-3}-\mathbf U_{N-2}
\right].
\end{equation}
The matrix $\boldsymbol{\Xi}$ has exactly $N-4$ block columns, each
containing $K$ columns.

We first show that $\boldsymbol{\Xi}$ has full column rank. Suppose that
vectors $\mathbf a_2,\mathbf a_3,\ldots,\mathbf a_{N-3}\in\mathbb F^K$
satisfy
\[
\sum_{h=2}^{N-3}
\left(
\mathbf U_h-\mathbf U_{h+1}
\right)\mathbf a_h
=
\mathbf 0.
\]
Using \eqref{eq:general-U-recurrence}, this relation becomes
\[
\left(
\mathbf I_{NT}-\boldsymbol{\Theta}
\right)
\sum_{h=2}^{N-3}\mathbf U_h\mathbf a_h
=
\mathbf 0.
\]
Since $\mathbf I_{NT}-\boldsymbol{\Theta}$ is invertible,
\[
\sum_{h=2}^{N-3}\mathbf U_h\mathbf a_h
=
\mathbf 0.
\]
Lemma~\ref{lemma:stacked-independence}, applied with $q=N-3$, gives
\[
\mathbf a_2=\mathbf a_3=\cdots=\mathbf a_{N-3}
=
\mathbf 0.
\]
Thus $\boldsymbol{\Xi}$ has full column rank, and hence
\begin{equation}
\label{eq:general-Xi-rank}
\operatorname{rank}(\boldsymbol{\Xi})
=
(N-4)K.
\end{equation}

We next identify the block rows of $\boldsymbol{\Xi}$. The $i$th
$T\times K$ block of $\mathbf U_h-\mathbf U_{h+1}$ is
\begin{align*}
\mathbf A_{i,h}\mathbf M_{i-h}
-
\mathbf A_{i,h+1}\mathbf M_{i-h-1}
&=
\mathbf A_{i,h}
\left(
\mathbf M_{i-h}
-
\boldsymbol{\Lambda}_{i-h}\mathbf M_{i-h-1}
\right)
=
\mathbf A_{i,h}\boldsymbol{\Phi}_{i-h}.
\end{align*}
Therefore, the $i$th $T\times (N-4)K$ block row of
$\boldsymbol{\Xi}$ is
\[
\left[
\mathbf A_{i,2}\boldsymbol{\Phi}_{i-2}
\;\;
\mathbf A_{i,3}\boldsymbol{\Phi}_{i-3}
\;\;
\cdots
\;\;
\mathbf A_{i,N-3}\boldsymbol{\Phi}_{i-N+3}
\right].
\]
By definition, the rank of this block row is $\mu_i$. Since $\boldsymbol{\Xi}$ is the vertical concatenation of these $N$ block
rows,
\[
\operatorname{rank}(\boldsymbol{\Xi})
\le
\sum_{i=0}^{N-1}\mu_i.
\]
Combining this inequality with \eqref{eq:general-Xi-rank} gives
\[
\sum_{i=0}^{N-1}\mu_i
\ge
(N-4)K. \qedhere
\]
\end{proof}

We can now complete the proof. Summing the local-overlap inequality
\eqref{eq:local_overlap} over all $i\in\mathcal V$ gives
\[
NK
\le
2Nd-\sum_{i=0}^{N-1}\mu_i.
\]
Applying Lemma~\ref{lemma:global-overlap}, we obtain
\[
NK
\le
2Nd-(N-4)K.
\]
Therefore
\[
d\ge\frac{N-2}{N}K.
\]
Since $d=T-K$,
\[
T
\ge
K+\frac{N-2}{N}K
=
\frac{2(N-1)}{N}K.
\]

This completes the proof for every $N\ge 5$, and hence completes the proof
of Theorem~\ref{theorem:necessary}.

%% file: Appendix.tex
\section{Proof of Theorem~\ref{theorem:necsuf}}
\label{appendix:necsuf}

    By \eqref{equ:y}, \eqref{equ:s}, and \eqref{eq:shuffle}, the complete
    received vector at node $i+1$ satisfies
    \begin{equation}\label{eq:s-T-shifted}
    \mathbf s_{i+1}^{(T)}
    =
    \mathbf M_i\mathbf x_i
    +
    \boldsymbol{\Lambda}_i\mathbf s_i^{(T)}.
    \end{equation}
    Here and throughout the proof, all node indices are reduced modulo $N$.

    Substituting \eqref{eq:s-T-shifted} successively around the ring exactly
    $N$ times gives
    \begin{IEEEeqnarray*}{rCl}
    \mathbf s_{i+1}^{(T)}
    =
    \mathbf M_i\mathbf x_i
    +
    \boldsymbol{\Lambda}_i\mathbf M_{i-1}\mathbf x_{i-1}
    +
    \boldsymbol{\Lambda}_i\boldsymbol{\Lambda}_{i-1}
    \mathbf M_{i-2}\mathbf x_{i-2}
    +\cdots
    +
    \mathbf A_{i,N-1}\mathbf M_{i+1}\mathbf x_{i+1}
    +
    \mathbf A_{i,N}\mathbf s_{i+1}^{(T)}.
    \end{IEEEeqnarray*}
    Using the notation $\mathbf A_{i,h}$, this can be written compactly as
    \[
    \mathbf s_{i+1}^{(T)}
    =
    \sum_{h=0}^{N-1}
    \mathbf A_{i,h}\mathbf M_{i-h}\mathbf x_{i-h}
    +
    \mathbf A_{i,N}\mathbf s_{i+1}^{(T)}.
    \]
    Equivalently,
    \[
    \left(
    \mathbf I_T-\mathbf A_{i,N}
    \right)
    \mathbf s_{i+1}^{(T)}
    =
    \sum_{h=0}^{N-1}
    \mathbf A_{i,h}\mathbf M_{i-h}\mathbf x_{i-h}.
    \]
    Since $\mathbf A_{i,N}$ is strictly lower triangular,
    $\mathbf I_T-\mathbf A_{i,N}$
    is invertible. Hence
    \[
    \mathbf s_{i+1}^{(T)}
    =
    \left(
    \mathbf I_T-\mathbf A_{i,N}
    \right)^{-1}
    \sum_{h=0}^{N-1}
    \mathbf A_{i,h}\mathbf M_{i-h}\mathbf x_{i-h}.
    \]

    The decoder at destination node $i+1$ outputs
    \begin{align*}
    \hat{\mathbf z}_{i+1}
    &=
    \mathbf R_{i+1}\mathbf s_{i+1}^{(T)}
    +
    \mathbf D_{i+1}\mathbf x_{i+1}
    \\
    &=
    \mathbf R_{i+1}
    \left(
    \mathbf I_T-\mathbf A_{i,N}
    \right)^{-1}
    \sum_{h=0}^{N-1}
    \mathbf A_{i,h}\mathbf M_{i-h}\mathbf x_{i-h}
    +
    \mathbf D_{i+1}\mathbf x_{i+1}
    \\
    &=
    \mathbf R'_{i+1}
    \sum_{h=0}^{N-1}
    \mathbf A_{i,h}\mathbf M_{i-h}\mathbf x_{i-h}
    +
    \mathbf D_{i+1}\mathbf x_{i+1},
    \end{align*}
    where 
    \[
    \mathbf R'_{i+1}
    =
    \mathbf R_{i+1}
    \left(
    \mathbf I_T-\mathbf A_{i,N}
    \right)^{-1}.
    \]

    Suppose first that the code is feasible. Then, for every destination node
    $i+1$ and every independently chosen collection of source vectors $\{\mathbf x_j\}_{j\in\mathcal V}$,
    \[
    \hat{\mathbf z}_{i+1}
    =
    \sum_{j=0}^{N-1}\mathbf x_j.
    \]
    Because the source vectors can be varied independently, the coefficient
    matrix multiplying each $\mathbf x_j$ must equal $\mathbf I_K$.

    For $h=0,1,\ldots,N-2$, the vectors
    \[
    \mathbf x_i,\mathbf x_{i-1},\ldots,\mathbf x_{i-N+2}
    \]
    are precisely the $N-1$ nonlocal source vectors at destination $i+1$.
    Their coefficient matrices are
    $\mathbf R'_{i+1}
    \mathbf A_{i,h}\mathbf M_{i-h}$,
    and hence feasibility requires
    \[
    \mathbf R'_{i+1}
    \mathbf A_{i,h}\mathbf M_{i-h}
    =
    \mathbf I_K,
    \qquad h=0,1,\ldots,N-2.
    \]
    This proves \eqref{equ:identity}.

    The remaining term, corresponding to $h=N-1$, contains the local source
    vector because
    $i-(N-1)\equiv i+1 \pmod N$.
    The total coefficient of $\mathbf x_{i+1}$ is therefore
    \[
    \mathbf R'_{i+1}
    \mathbf A_{i,N-1}\mathbf M_{i+1}
    +
    \mathbf D_{i+1}.
    \]
    Feasibility requires this coefficient to equal $\mathbf I_K$, which gives
    \eqref{equ:txi}. This proves necessity.

    Conversely, suppose that \eqref{equ:identity} and \eqref{equ:txi} hold.
    Substituting these identities into the decoder expression gives
    $\hat{\mathbf z}_{i+1}
    =
    \sum_{h=0}^{N-1}\mathbf x_{i-h}$
    for every destination and every choice of source vectors. Therefore the
    linear network code is feasible.

\section{Time-Unrolled Feasibility Characterization}
\label{section:TUFC}

The feasibility conditions in Theorem~\ref{theorem:necsuf} admit an
equivalent form that explicitly accounts for every possible number of
traversals around the ring.

Recall that, for every $\ell\ge 0$,
\[
\mathbf A_{i,0}=\mathbf I_T,
\qquad
\mathbf A_{i,\ell}
=
\boldsymbol{\Lambda}_i
\boldsymbol{\Lambda}_{i-1}
\cdots
\boldsymbol{\Lambda}_{i-\ell+1},
\]
where all node indices are reduced modulo $N$. Because each
$\boldsymbol{\Lambda}_j$ is strictly lower triangular, any product of at
least $T$ such matrices is zero. In particular,
\begin{equation}
\label{eq:A-zero-long-path}
\mathbf A_{i,\ell}=\mathbf 0,
\qquad \ell\ge T.
\end{equation}

Moreover, periodicity of the node indices gives
\begin{equation}
\label{eq:A-cycle-factorization}
\mathbf A_{i,mN+h}
=
\left(\mathbf A_{i,N}\right)^m\mathbf A_{i,h},
\qquad
m\ge 0,\quad 0\le h<N.
\end{equation}

Since $\mathbf A_{i,N}$ is strictly lower triangular, we have 
\[
\left(\mathbf I_T-\mathbf A_{i,N}\right)^{-1}
=
\sum_{m=0}^{T-1}\left(\mathbf A_{i,N}\right)^m.
\]
Combining this finite Neumann expansion with
\eqref{eq:A-zero-long-path} and \eqref{eq:A-cycle-factorization} gives
\begin{equation}
\label{eq:neumann-residue-expansion}
\left(\mathbf I_T-\mathbf A_{i,N}\right)^{-1}
\mathbf A_{i,h}
=
\sum_{\substack{m\ge 0\\mN+h<T}}
\mathbf A_{i,mN+h},
\qquad 0\le h<N.
\end{equation}

Write the Euclidean division of $T$ by $N$ as
\[
T=aN+b,
\qquad
a\ge 0,\quad 0\le b<N.
\]
For each $h\in\{0,1,\ldots,N-1\}$, define
\[
\eta_h
=
a+\mathbf 1_{\{h<b\}},
\]
where $\mathbf 1_{\{h<b\}}$ is equal to $1$ when $h<b$ and to $0$
otherwise. Equivalently,
\[
\eta_h
=
\left|
\left\{
m\in\mathbb Z_{\ge 0}:mN+h\le T-1
\right\}
\right|.
\]
Thus \eqref{eq:neumann-residue-expansion} becomes
\begin{equation}
\label{eq:neumann-eta-expansion}
\left(\mathbf I_T-\mathbf A_{i,N}\right)^{-1}
\mathbf A_{i,h}
=
\sum_{m=0}^{\eta_h-1}\mathbf A_{i,mN+h}.
\end{equation}
Throughout this section, a sum with an empty index set is understood to
be the zero matrix.

\begin{theorem}[Time-unrolled feasibility characterization]
\label{theorem:necsuf-2}
A $(K,T)$ linear network code for the All-Reduce problem over the $N$-node
directed ring is feasible if and only if
\begin{align}
\mathbf R_{i+1}
\left(
\sum_{m=0}^{\eta_h-1}
\mathbf A_{i,mN+h}
\right)
\mathbf M_{i-h}
&=
\mathbf I_K,\label{equ:identity-expanded} \quad \forall i\in\mathcal V,
h=0,1,\ldots,N-2,
\\
\mathbf R_{i+1}
\left(
\sum_{m=0}^{a-1}
\mathbf A_{i,mN+N-1}
\right)
\mathbf M_{i+1}
+
\mathbf D_{i+1}
&=
\mathbf I_K, \quad 
\forall i\in\mathcal V.
\label{equ:txi-expanded}
\end{align}
\end{theorem}

\begin{remark}
The matrices
\[
\sum_{m=0}^{\eta_h-1}\mathbf A_{i,mN+h}
\]
collect all causal propagation paths from source node $i-h$ to destination
node $i+1$ whose lengths are congruent to $h$ modulo $N$. The index $m$
counts the number of complete additional traversals around the ring.

As in Theorem~\ref{theorem:necsuf}, the substantive constraints are
\eqref{equ:identity-expanded}. Once these constraints are satisfied, the
local-decoding matrices can be chosen as
\[
\mathbf D_{i+1}
=
\mathbf I_K
-
\mathbf R_{i+1}
\left(
\sum_{m=0}^{a-1}
\mathbf A_{i,mN+N-1}
\right)
\mathbf M_{i+1}.
\]
\end{remark}

\begin{proof}
From the one-cycle expansion established in the proof of
Theorem~\ref{theorem:necsuf}, the received vector at destination node
$i+1$ is
\begin{equation}
\label{eq:s-one-cycle-inverse}
\mathbf s_{i+1}^{(T)}
=
\left(\mathbf I_T-\mathbf A_{i,N}\right)^{-1}
\sum_{h=0}^{N-1}
\mathbf A_{i,h}\mathbf M_{i-h}\mathbf x_{i-h}.
\end{equation}
Applying \eqref{eq:neumann-eta-expansion} to each term in
\eqref{eq:s-one-cycle-inverse} gives
\begin{equation}
\label{eq:s-expanded-shifted}
\mathbf s_{i+1}^{(T)}
=
\sum_{h=0}^{N-1}
\left(
\sum_{m=0}^{\eta_h-1}
\mathbf A_{i,mN+h}
\right)
\mathbf M_{i-h}\mathbf x_{i-h}.
\end{equation}

Equation \eqref{eq:s-expanded-shifted} is the direct time-unrolled form of
the received vector. Indeed, every integer
$r\in\{0,1,\ldots,T-1\}$ has a unique representation
\[
r=mN+h,
\qquad
0\le h<N,
\]
and the term with index $r=mN+h$ represents a propagation path of length
$r$ from source node $i-h$ to destination node $i+1$.

The decoder at node $i+1$ produces
\[
\hat{\mathbf z}_{i+1}
=
\mathbf R_{i+1}\mathbf s_{i+1}^{(T)}
+
\mathbf D_{i+1}\mathbf x_{i+1}.
\]
Substituting \eqref{eq:s-expanded-shifted} yields
\begin{IEEEeqnarray}{rCl}
\hat{\mathbf z}_{i+1}
=
\sum_{h=0}^{N-1}
\mathbf R_{i+1}
\left(
\sum_{m=0}^{\eta_h-1}
\mathbf A_{i,mN+h}
\right)
\mathbf M_{i-h}\mathbf x_{i-h}
+
\mathbf D_{i+1}\mathbf x_{i+1}.
\label{eq:decoder-expanded}
\end{IEEEeqnarray}

Suppose first that the code is feasible. Then
\[
\hat{\mathbf z}_{i+1}
=
\sum_{j\in\mathcal V}\mathbf x_j
\]
for every destination node and every independently chosen collection of
source vectors. Consequently, the coefficient matrix multiplying each
source vector in \eqref{eq:decoder-expanded} must be $\mathbf I_K$.

For $h=0,1,\ldots,N-2$, the vectors
\[
\mathbf x_i,\mathbf x_{i-1},\ldots,\mathbf x_{i-N+2}
\]
are precisely the nonlocal source vectors at destination $i+1$.
Their coefficient matrices therefore satisfy
\[
\mathbf R_{i+1}
\left(
\sum_{m=0}^{\eta_h-1}
\mathbf A_{i,mN+h}
\right)
\mathbf M_{i-h}
=
\mathbf I_K,
\]
which gives \eqref{equ:identity-expanded}.

For $h=N-1$, the corresponding source is local because
$i-(N-1)\equiv i+1\pmod N$.
Since $0\le b<N$, we have
\[
\eta_{N-1}
=
a+\mathbf 1_{\{N-1<b\}}
=
a.
\]
Thus the total coefficient of $\mathbf x_{i+1}$ is
\[
\mathbf R_{i+1}
\left(
\sum_{m=0}^{a-1}
\mathbf A_{i,mN+N-1}
\right)
\mathbf M_{i+1}
+
\mathbf D_{i+1}.
\]
Equating this coefficient to $\mathbf I_K$ gives
\eqref{equ:txi-expanded}. This proves necessity.

Conversely, suppose that \eqref{equ:identity-expanded} and
\eqref{equ:txi-expanded} hold. Substituting these identities into
\eqref{eq:decoder-expanded} gives
\[
\hat{\mathbf z}_{i+1}
=
\sum_{h=0}^{N-2}\mathbf x_{i-h}
+
\mathbf x_{i+1}
=
\sum_{j\in\mathcal V}\mathbf x_j.
\]
This holds for every $i\in\mathcal V$ and every collection of source
vectors. Hence every destination recovers the desired sum, and the code is
feasible.
\end{proof}

\section{Contiguous Transmissions in the Interleaved Schedule}
\label{section:RMFC}

The following lemmas are used in the proof of Theorem~\ref{theorem:interleaved}.
Fix $N\ge 3$ and $1\leq r<N$. Recall that $\Upsilon=\{\lfloor jN/r\rfloor:0\le j<r\}$ is the set of nodes that inject an additional coordinate in time unit \(1\). 

\begin{lemma}
\label{lemma:interleaved-contiguous}
In the interleaved reduce--multicast code for $N+r$ coordinates, each node transmits in consecutive time units starting at time unit \(0\).
Once a node is idle, it never transmits again.
\end{lemma}

\begin{proof}
All node indices are taken modulo \(N\). Write \((a)^+=\max\{a,0\}\). For each \(v\in \Upsilon\), regard the
coordinate that node \(v\) injects in time unit \(1\) as already held in
reserve in time unit \(0\), with \(2(N-1)\) transmissions remaining. Thus, a
node has at most two coordinates available in any time unit: one retained
from the previous time unit and one newly received from its predecessor.
It transmits one of the available coordinates and retains the other,
if any.

Let \(\alpha_t(v)\) and \(\beta_t(v)\) be the remaining-transmission counts of the
coordinate sent and retained by node \(v\) in time unit \(t\), respectively. Set \(\alpha_t(v)=0\) if \(v\) is idle and \(\beta_t(v)=0\) if none is retained.
In time unit \(0\), \(\alpha_0(v)=2(N-1)\), while \(\beta_0(v)=0\) for $v\notin\Upsilon$ and \(\beta_0(v)=2(N-1)\) for $v\in\Upsilon$.

A coordinate sent by node \(v-1\) in time unit \(t\) arrives at node \(v\)
with \((\alpha_t(v-1)-1)^+\) transmissions remaining. The priority rule
therefore gives
\begin{align}
\alpha_{t+1}(v)
=\max\!\left\{\beta_t(v),(\alpha_t(v-1)-1)^+\right\}, \quad
\beta_{t+1}(v)
=\min\!\left\{\beta_t(v),(\alpha_t(v-1)-1)^+\right\}.
\label{eq:interleaved-registers}
\end{align}
These counts are
unaffected by how ties are broken. Node \(v\) transmits in time unit
\(t\) exactly when \(\alpha_t(v)>0\).

We show by induction that \(\alpha_{t+1}(v)\leq \alpha_t(v)\) for every \(t\) and \(v\).
The claim holds at \(t=0\), since no count exceeds \(2(N-1)\). If it holds
at time \(t\), then  \eqref{eq:interleaved-registers} gives
\begin{align*}
   \beta_{t+1}(v)\leq \beta_t(v)\leq \alpha_{t+1}(v),\quad
(\alpha_{t+1}(v-1)-1)^+
\leq (\alpha_t(v-1)-1)^+
\leq \alpha_{t+1}(v). 
\end{align*}
Hence, 
\[\alpha_{t+2}(v)
=\max\!\left\{\beta_{t+1}(v),(\alpha_{t+1}(v-1)-1)^+\right\}\leq \alpha_{t+1}(v),\]
completing the induction. 
Therefore, \(\alpha_t(v)\) is nonincreasing in \(t\). If it is positive in a
given time unit, it was positive in every earlier time unit. Once it reaches
zero, it remains zero. The stated transmission pattern follows.
\end{proof}

\begin{lemma} \label{lemma:Lv-max}
    For $1\leq r<N$,
   the transmission counts in
\eqref{eq:total-transmissions} satisfy
\[
\max_{v\in\mathcal V}L_v
=
2(N+r-1)-\left\lfloor\frac{2r}{N}\right\rfloor.
\]
\end{lemma}

\begin{proof}
By \eqref{eq:total-transmissions},
$L_v=2(N+r-1)
-\mathbf 1[v+1\in \Upsilon]-\mathbf 1[v+2\in\Upsilon]$. It suffices to find the minimum, over \(v\), of the sum of the two indicators. Let
\[a_v=\mathbf 1[v+1\in\Upsilon]+\mathbf 1[v+2\in \Upsilon].\]
Since $\sum_{v\in\mathcal V}a_v=2r$, if $2r<N$, some $a_v=0$. If $2r\geq N$, the selected nodes in \(\Upsilon\) leave no two consecutive nodes unselected, so $a_v\geq 1$ for every $v$; because $r<N$, some $a_v=1$. Thus, in both cases
\[\min_{v\in\mathcal V}a_v=\left\lfloor\frac{2r}{N}\right\rfloor.
\]
This completes the proof.
\end{proof}

\section{Proof of Theorem~\ref{thm:k-less-n}}
\label{section:proof-k-less-n}
Sufficiency follows from the reduce--partial-multicast construction. We prove necessity. 

Suppose a feasible coordinate-wise linear network code completes in
\(T=N+K-2\) time units. Time units are indexed by
\(0,\ldots,T-1\).
All node indices are taken modulo
\(N\).

A \emph{dependency path} from a source symbol \(x_{i,c}\) to a
destination consists of a sequence of transmissions
\[
y_{v_0\rightarrow}^{(t_0)},\
y_{v_1\rightarrow}^{(t_1)},\ \ldots,\
y_{v_{\ell-1}\rightarrow}^{(t_{\ell-1})},
\]
where \(v_0=i\), \(v_{a+1}=v_a+1\), and
\(t_0<t_1<\cdots<t_{\ell-1}\). The first transmission includes
\(x_{i,c}\) with a nonzero encoding coefficient. Each subsequent
transmission includes the symbol received from the preceding
transmission with a nonzero recoding coefficient, possibly after
storage. The final transmission reaches the destination and its
received symbol is used with a nonzero coefficient in decoding
coordinate \(c\).
Expanding the encoding, recoding, and decoding operations expresses
the coefficient of \(x_{i,c}\) in the decoded output as a sum of
products along dependency paths. Since feasibility requires this
coefficient to equal \(1\), at least one such path has a nonzero
product. In a coordinate-wise code, every transmission on that path
is associated with coordinate \(c\).

Fix a source \(i\) and a coordinate \(c\). Feasibility requires node
\(i-1\) to recover the coordinate sum with coefficient \(1\) on
\(x_{i,c}\). Hence, a dependency path exists from \(x_{i,c}\) to
node \(i-1\).
The path starts at node \(i\) and follows successive edges of the
directed ring. It therefore contains at least \(N-1\) transmissions.
For \(d=0,\ldots,N-2\), let \(t_d\) be the time unit of its
\((d+1)\)-st transmission, which goes from node \(i+d\) to node
\(i+d+1\). 
There are \(d\) transmissions before this one. Since successive
transmissions along a dependency path occur in strictly increasing
time units, we have \(t_d\geq d\). After this transmission, at least
\(N-2-d\) further transmissions are needed to reach node \(i-1\).
All must occur by time unit \(T-1\), so
$t_d+(N-2-d)\leq T-1$.
Using \(T=N+K-2\), we obtain
$d\leq t_d\leq d+K-1$.
Thus, the \((d+1)\)-st transmission occurs in the window
\[
W_d=\{d,\ldots,d+K-1\}.
\]
Applying this argument to all \(K\) coordinates shows that the edge
\(i+d\to i+d+1\) must transmit every coordinate within \(W_d\).
Since the window contains exactly \(K\) time units and each edge transmits
at most one coordinate per time unit, every coordinate appears exactly
once, with no idle time units.

Now fix an edge \(j\to j+1\). For each \(d=0,\ldots,N-2\), choosing
source \(i=j-d\) shows that this edge transmits every coordinate
exactly once in \(W_d\). Consecutive windows \(W_d\) and \(W_{d+1}\)
differ only in time units \(d\) and \(d+K\). Because both contain every
coordinate exactly once, the coordinates transmitted in these two
time units must be the same. Hence, the coordinate schedule on each edge
is \(K\)-periodic throughout the execution.

Let \(p_j(c)\in\{0,\ldots,K-1\}\) denote the time unit in which edge
\(j\to j+1\) transmits coordinate \(c\) in \(W_0\). For each \(j\),
the values \(p_j(c)\), \(c=0,\ldots,K-1\), form a permutation of
\(\{0,\ldots,K-1\}\).
Return to the dependency path from \(x_{i,c}\) to node \(i-1\).
Its \((d+1)\)-st transmission must be the unique coordinate-\(c\)
transmission on edge \(i+d\to i+d+1\) within $W_d$.
Let \(t_d(c)\) denote its time unit. By periodicity,
\[
t_d(c)\equiv p_{i+d}(c)\pmod K.
\]
Write $t_d(c)=d+o_d(c)$,
where \(0\leq o_d(c)\leq K-1\) is the offset from the start of \(W_d\). It follows
that
\begin{equation}
\label{equ:residual}
    o_d(c)=\bigl(p_{i+d}(c)-d\bigr)\bmod K.
\end{equation}

Successive transmissions along the dependency path occur in strictly
increasing time units. Therefore,
\[
t_{d+1}(c)\geq t_d(c)+1,\quad d=0,\ldots,N-3,
\]
which, after substituting \(t_d(c)=d+o_d(c)\), gives
\begin{equation}
\label{equ:oo}
    o_{d+1}(c)\geq o_d(c).
\end{equation}

For each \(d\), the values \(p_{i+d}(c), c=0,\ldots,K-1\) form a permutation of
\(\{0,\ldots,K-1\}\). By \eqref{equ:residual}, \(o_d(c)\) also form a permutation. Hence,
\begin{equation}
\label{equ:sum}
   \sum_{c=0}^{K-1}o_d(c)=\frac{K(K-1)}{2}. 
\end{equation}

\eqref{equ:oo} and \eqref{equ:sum} imply
\[
o_{d+1}(c)=o_d(c)
\]
for every coordinate \(c\). In particular, taking \(d=0\), which is
permitted because \(N\geq3\), yields
\[
p_{i+1}(c)\equiv p_i(c)+1\pmod K.
\]
This relation holds for every node \(i\). Iterating it around the ring
gives
\[
p_i(c)\equiv p_i(c)+N\pmod K.
\]
Therefore, \(K\) divides \(N\), proving necessity. This completes the proof of Theorem~\ref{thm:k-less-n}.

\section{Invertibility of the Decoding Matrices}
\label{app:decoding-invertibility}

\subsection{Properties of the Recursive Corner Block}
\begin{lemma}[Submatrices of the recursive corner block]
\label{lem:H-properties}
Let \(\mathbf H(a,b)\) be the recursively defined corner block, with
\(a,b\geq1\) in \eqref{equ:H}. Then:
\begin{enumerate}
    \item Every leading square submatrix of order
    \(1\leq \ell\leq\min(a,b)\) has determinant \(\pm1\).

    \item If \(a\leq b\), every submatrix formed by \(a\)
    consecutive columns has determinant \(\pm1\).

    \item If \(a\geq b\), every submatrix formed by \(b\)
    consecutive rows has determinant \(\pm1\).

    \item Every trailing square submatrix of order
    \(1\leq \ell\leq\min(a,b)\) equals \(\mathbf I_\ell\).
\end{enumerate}
Here, consecutive indices are taken in their ordinary order,
without wraparound. All determinants are evaluated over the
integers. Consequently, the submatrices in the first three
statements are invertible over every field.
\end{lemma}

\begin{proof}

We prove all four statements together by strong induction on \(a+b\).
If \(a=b\), then \(\mathbf H(a,b)=\mathbf I_a\), and the statements follow
immediately. If \(a\neq b\), assume that all four statements hold for
every pair with a smaller dimension sum. The matrix \(\mathbf H(a,b)\)
is obtained by appending an identity block to either \(\mathbf H(a-b,b)\)
or \(\mathbf H(a,b-a)\). These smaller matrices have dimension sums \(a\)
and \(b\), respectively, so we may use all four properties for the
recursive block. We show below that the properties remain valid
after the identity block is appended.

The fourth statement follows from the appended identity block:
if \(a>b\), the last \(b\) rows form \(\mathbf I_b\); if
\(a<b\), the last \(a\) columns form \(\mathbf I_a\).

\paragraph{Leading square submatrices.}
Suppose first that \(a>b\). Then
\begin{equation}
\label{equ:Hab1}
    \mathbf H(a,b)
=
\begin{bmatrix}
H(a-b,b)\\
\mathbf I_b
\end{bmatrix}.
\end{equation}
For \(\ell\leq a-b\), the leading submatrix of order \(\ell\) lies
entirely in \(\mathbf H(a-b,b)\), so induction applies. For \(\ell>a-b\), set \(d=\ell-(a-b)\). The bottom \(d\) rows of the
leading submatrix are
\(\begin{bmatrix}\mathbf I_d&\mathbf 0\end{bmatrix}\).
Moving its first \(d\) columns to the end gives
\[
\begin{bmatrix}
\mathbf C & \mathbf X\\
\mathbf 0 & \mathbf I_d
\end{bmatrix},
\qquad
\mathbf C=\mathbf H(a-b,b)[\,0:a-b-1,\ d:\ell-1\,],
\]
where \(\mathbf X\) contains the remaining entries. Throughout the proof, \(\mathbf X\) denotes a block whose entries
do not affect the determinant; its dimensions are determined by
the surrounding blocks.
The matrix \(\mathbf C\) consists of \(a-b\) consecutive columns
of \(\mathbf H(a-b,b)\). Since \(a-b<\ell\leq b\), statement 2 of the
induction hypothesis gives \(\det(\mathbf C)=\pm1\).
The column permutation changes only the sign.

Now suppose \(a<b\). Then
\begin{equation}
\label{equ:Hab}
    \mathbf H(a,b)
=
\begin{bmatrix}
\mathbf H(a,b-a)&\mathbf I_a
\end{bmatrix}.
\end{equation}
For \(\ell\leq b-a\), induction applies directly. For \(\ell>b-a\), set
\(d=\ell-(b-a)\). The last \(d\) columns of the leading submatrix
are
\(\begin{bmatrix}\mathbf I_d\\\mathbf 0\end{bmatrix}\).
Moving its first \(d\) rows to the bottom gives
\[
\begin{bmatrix}
\mathbf C & \mathbf 0\\
\mathbf X & \mathbf I_d
\end{bmatrix},
\qquad
\mathbf C=\mathbf H(a,b-a)[\,d:\ell-1,\ 0:b-a-1\,].
\]
Here, \(\mathbf C\) consists of \(b-a\) consecutive rows of
\(\mathbf H(a,b-a)\). Since \(b-a<\ell\leq a\), statement 3 of the
induction hypothesis gives \(\det(\mathbf C)=\pm1\).
Thus, statement 1 holds in both cases.

\paragraph{Consecutive-column submatrices.}
Suppose \(a<b\), then $\mathbf H(a,b)$ is given in the form of \eqref{equ:Hab}.
Consider a window of \(a\) consecutive columns. If it lies
entirely in \(\mathbf H(a,b-a)\), induction applies. If it lies
entirely in \(\mathbf I_a\), it is \(\mathbf I_a\).
Otherwise, the window contains the last \(k\) columns of
\(\mathbf H(a,b-a)\) followed by the first \(a-k\) columns of
\(\mathbf I_a\), where
\(1\leq k<a\) and \(k\leq b-a\).
By statement 4 of the induction hypothesis, the bottom \(k\)
rows of these last \(k\) columns of \(\mathbf H(a,b-a)\) form
\(\mathbf I_k\). Moving those rows to the top gives
\[
\begin{bmatrix}
\mathbf I_k & \mathbf 0\\
\mathbf X & \mathbf I_{a-k}
\end{bmatrix}.
\]
Its determinant is \(1\), so the original window has determinant
\(\pm1\). The case \(a=b\) is covered by the base case.

\paragraph{Consecutive-row submatrices.}
Suppose \(a>b\), and then $\mathbf H(a,b)$ is given in the form of \eqref{equ:Hab1}.
A window of \(b\) consecutive rows lying entirely in
\(\mathbf H(a-b,b)\) satisfies the claim by induction; one lying
entirely in \(\mathbf I_b\) is \(\mathbf I_b\).
A window straddling the two blocks contains the last \(k\)
rows of \(\mathbf H(a-b,b)\) and the first \(b-k\) rows of
\(\mathbf I_b\), where
\(1\leq k<b\) and \(k\leq a-b
\).
By statement 4 of the induction hypothesis, the last \(k\)
columns of those rows of \(\mathbf H(a-b,b)\) form \(\mathbf I_k\).
Moving these columns to the left gives
\[
\begin{bmatrix}
\mathbf I_k & \mathbf X\\
\mathbf 0 & \mathbf I_{b-k}
\end{bmatrix}.
\]
Its determinant is \(1\), so the original window has determinant
\(\pm1\). This completes the induction.
\end{proof}

\subsection{Determinants of the Submatrices of $\mathbf W$}

\begin{lemma}
\label{lem:determinant}
Let \(\mathbf W\in\mathbb F^{\varrho\times K}\) be the matrix defined in Construction \ref{def:joint-linear-construction}. Then \(\mathbf W[0,\varrho]=1\), and each of the
following square submatrices has determinant \(\pm1\):
\begin{enumerate}
    \item
    \(
    \mathbf W[\,0:i+1,\ \varrho:\varrho+i+1\,],
    \quad 0\leq i\leq\varrho-2;
    \)

    \item
    \(
    \mathbf W[\,0:\varrho-1,\ i+2:i+\varrho+1\,],
    \quad \varrho-1\leq i\leq K-2;
    \)

    \item
    \(
    \mathbf W[\,i-K+2:\varrho-1,\ i-K+2:\varrho-1\,],
    \quad K-1\leq i\leq K+\varrho-3.
    \)
\end{enumerate}
Here, \(\mathbf W[a:b,c:d]\) selects rows
\(a,a+1\ldots,b\) and columns \(c,\ldots,d\), preserving their listed
order. Column indices are taken modulo \(K\). Empty index ranges are omitted. Thus, all these submatrices are
invertible over every field.
\end{lemma}

\begin{proof}
Write
\[
K=q\varrho+\xi,
\qquad q\geq1,\quad 0\leq\xi<\varrho.
\]
All column indices of \(\mathbf W\) are taken modulo \(K\),
preserving their listed order.

If \(\xi=0\), then \(\mathbf W\) consists of \(q\) copies of
\(\mathbf I_\varrho\). The first and third families select identity
submatrices, while the second selects permutation matrices. Hence,
all the stated determinants are \(\pm1\), and
\(\mathbf W[0,\varrho]=1\).

Now suppose \(\xi>0\). The first
\(\varrho\) columns of \(\mathbf W\) form
\[
\mathbf C
=
\begin{bmatrix}
\mathbf I_\xi & \mathbf 0\\
\mathbf H(\varrho-\xi,\xi) & \mathbf I_{\varrho-\xi}
\end{bmatrix}.
\]
The remaining columns consist of \(q-1\) copies of
\(\mathbf I_\varrho\), followed by its first \(\xi\) columns.
Thus, \(\det(\mathbf C)=1\) and
\(\mathbf W[0,\varrho]=1\). We use
Lemma~\ref{lem:H-properties} to verify the three families below.
All row and column permutations change only the sign of a
determinant.

\setcounter{paragraph}{0}
\paragraph{First family.}
Consider
\[
\mathbf A_i
=
\mathbf W[\,0:i+1,\ \varrho:\varrho+i+1\,],
\qquad 0\leq i\leq\varrho-2.
\]
If \(q\geq2\), the selected columns lie in the second identity
block, so \(\mathbf A_i=\mathbf I_{i+2}\). If \(q=1\) and
\(i+2\leq\xi\), they lie in the final block of width \(\xi\),
and again \(\mathbf A_i=\mathbf I_{i+2}\).

It remains to consider \(q=1\) and \(i+2>\xi\). In this case,
\(K=\varrho+\xi\), and the selected columns wrap around \(K\).
Set \(L=i+2-\xi\), so \(1\leq L\leq \varrho-\xi\). 
\[
\mathbf A_i
=
\begin{bmatrix}
\mathbf I_\xi & \mathbf X\\
\mathbf 0 &
\mathbf C[\,\xi:\xi+L-1,\ 0:L-1\,]
\end{bmatrix},
\]
where \(\mathbf X\) contains the remaining entries. Throughout the proof, \(\mathbf X\) denotes a block whose entries
do not affect the determinant. By the block-triangular form,
\[
\det(\mathbf A_i)
=
\det\!\left(
\mathbf C[\,\xi:\xi+L-1,\ 0:L-1\,]
\right).
\]
If \(L\leq\xi\), \(\mathbf C[\,\xi:\xi+L-1,\ 0:L-1\,]\) is the leading square
submatrix of \(\mathbf H(\varrho-\xi,\xi)\) of order \(L\). Its
determinant is \(\pm1\) by Lemma~\ref{lem:H-properties}.
If \(L>\xi\), partition the  block as
\[
\mathbf C[\,\xi:\xi+L-1,\ 0:L-1\,]
=
\begin{bmatrix}
\mathbf X & \mathbf I_{L-\xi}\\
\mathbf D & \mathbf 0
\end{bmatrix},
\qquad
\mathbf D
=
\mathbf H(\varrho-\xi,\xi)[\,L-\xi:L-1,\ 0:\xi-1\,].
\]
Moving the last \(L-\xi\) columns to the front gives
\[
\begin{bmatrix}
\mathbf I_{L-\xi} & \mathbf X\\
\mathbf 0 & \mathbf D
\end{bmatrix}.
\]
Hence, \(\det(\mathbf C[\,\xi:\xi+L-1,\ 0:L-1\,])=\pm\det(\mathbf D)\). 
Since \(\xi<L\leq\varrho-\xi\), the matrix
\(\mathbf H(\varrho-\xi,\xi)\) has more rows than columns.
\(\mathbf D\) consists of \(\xi\) consecutive rows,
so Lemma~\ref{lem:H-properties} gives
\(\det(\mathbf D)=\pm1\). Therefore,
\(\det(\mathbf A_i)=\pm1\).

\paragraph{Second family.}
Consider
\[
\mathbf B_i
=
\mathbf W[\,0:\varrho-1,\ i+2:i+\varrho+1\,],
\qquad \varrho-1\leq i\leq K-2.
\]
If the selected columns do not wrap around \(K\), they are
unmodified standard basis columns with distinct residues modulo
\(\varrho\). Thus, \(\mathbf B_i\) is a permutation matrix.
If \(i=K-2\), then \(\mathbf B_i=\mathbf C\), whose determinant
is \(1\).

In the remaining cases, the selected columns wrap around \(K\),
and
\[
\mathbf B_i
=
\begin{bmatrix}
\mathbf W[\,0:\varrho-1,\ i+2:K-1\,]
&
\mathbf C[\,0:\varrho-1,\ 0:i+\varrho+1-K\,]
\end{bmatrix}.
\]
Set \(L=K-i-2\). We consider the following cases.

\emph{Case 1: \(L\leq\xi\).}
The first \(L\) columns of \(\mathbf B_i\) are standard basis columns selecting
rows \(\xi-L,\ldots,\xi-1\).
Move these rows to the top, preserving the order of the selected
columns. The matrix then has the form
\[
\begin{bmatrix}
\mathbf I_{L} & \mathbf X\\
\mathbf 0 & \mathbf D
\end{bmatrix},
\]
where \(\mathbf D\) consists of the first
\(\varrho-L\) columns of \(\mathbf C\), with rows
\(\xi-L,\ldots,\xi-1\) removed. Hence,
\(\det(\mathbf B_i)=\pm\det(\mathbf D)\).

If \(L\leq\varrho-\xi\), the retained identity blocks in \(\mathbf D\) have orders
\(\xi-L\) and \(\varrho-\xi-L\). Using the first identity block in \(\mathbf D\)
to eliminate the entries below it, then permuting columns, gives
\[
\begin{bmatrix}
\mathbf I_{\xi-L} & \mathbf0 & \mathbf0\\
\mathbf0 & \mathbf I_{\varrho-\xi-L} & \mathbf X\\
\mathbf0 & \mathbf0 & \mathbf E
\end{bmatrix},
\]
where
\(\mathbf E=
\mathbf H(\varrho-\xi,\xi)
[\,\varrho-\xi-L:\varrho-\xi-1,\,
  \xi-L:\xi-1\,]\), the trailing square submatrix of
order $L$. 
Thus,  by
Lemma~\ref{lem:H-properties}, \(\det(\mathbf D)=\pm\det(\mathbf E)=\pm1\).

If \(\varrho-\xi<L\leq\xi\), the retained identity block in \(\mathbf D\) have order
\(\xi-L\).
After placing these columns first, we obtain
\[
\begin{bmatrix}
\mathbf I_{\xi-L} & \mathbf X\\
\mathbf 0 & \mathbf E
\end{bmatrix},
\]
where
\[
\mathbf E
=
\mathbf H(\varrho-\xi,\xi)
[\,0:\varrho-\xi-1,\,
  \xi-L:\varrho-L-1\,].
\]
This matrix consists of \(\varrho-\xi\) consecutive columns of
\(\mathbf H(\varrho-\xi,\xi)\). Therefore,
\(\det(\mathbf D)=\pm1\) by Lemma~\ref{lem:H-properties}.
Hence, \(\det(\mathbf B_i)=\pm1\) throughout Case 1.

\emph{Case 2: \(L>\xi\).}
The first \(L\) columns of \(\mathbf B_i\) select rows
\[
0,\ldots,\xi-1
\quad\text{and}\quad
\varrho+\xi-L,\ldots,\varrho-1.
\]
Move these rows to the top, in the order matching the selected
columns. The resulting matrix is
\[
\begin{bmatrix}
\mathbf I_{L} & \mathbf X\\
\mathbf 0 &
\mathbf C[\,\xi:\xi+\varrho-L-1,\ 0:\varrho-L-1\,]
\end{bmatrix}.
\]
Consequently,
\[
\det(\mathbf B_i)
=
\pm\det\!\left(
\mathbf C[\,\xi:\xi+\varrho-L-1,\ 0:\varrho-L-1\,]
\right).
\]
Here, \(1\leq\varrho-L <\varrho-\xi\). This remaining block is exactly
the type considered in the first family. If \(L\geq\varrho-\xi\), it is
a leading square submatrix of \(\mathbf H(\varrho-\xi,\xi)\).
If \(L<\varrho-\xi\), it has the form
\[
\begin{bmatrix}
\mathbf X & \mathbf I_{\varrho-L-\xi}\\
\mathbf D & \mathbf 0
\end{bmatrix},
\qquad
\mathbf D
=
\mathbf H(\varrho-\xi,\xi)
[\,\varrho-L-\xi:\varrho-L-1,\ 0:\xi-1\,].
\]
Moving the last \(\varrho-L-\xi\) columns to the front gives
\[
\begin{bmatrix}
\mathbf I_{\varrho-L-\xi} & \mathbf X\\
\mathbf 0 & \mathbf D
\end{bmatrix}.
\]
Lemma~\ref{lem:H-properties} gives determinant \(\pm1\) in
both cases. Therefore, every submatrix in the second family
has determinant \(\pm1\).

\paragraph{Third family.}
Each selected submatrix is a trailing principal submatrix of
\(\mathbf C\), which is block lower triangular with identity
diagonal blocks. Its determinant is therefore \(1\).

This proves all the stated determinant identities. Since
\(\pm1\) is nonzero over every field, all the selected
submatrices are invertible.
\end{proof}

\subsection{Proof of Lemma~\ref{lem:Qi-invertibility}}
\begin{proof}
All node indices are taken modulo \(N\). The rows of \(\mathbf Q_i\)
correspond to the \(K\) consecutive starter nodes
\[
i+1,\ i,\ i-1,\ \ldots,\ i-K+2.
\]
We first consider only the rows contributed by starter nodes
\(j\geq\varrho\), each of which equals
\(\mathbf e_{j\bmod K}\). We show that these rows select distinct
coordinates within each \(\mathbf Q_i\).

For \(K-2\leq i\leq N-2\), the starter indices form a consecutive
interval of length \(K\), with distinct residues modulo \(K\).
The subset of starters satisfying \(j\geq\varrho\) therefore also
selects distinct coordinates.

For \(0\leq i\leq K-3\), the starter list wraps below \(0\).
The starters satisfying \(j\geq\varrho\) lie in the two intervals
\[
\{N+i-K+2,\ldots,N-1\}
\quad\text{and}\quad
\{\varrho,\ldots,i+1\},
\]
where an empty interval is omitted. Since \(N\bmod K=\varrho\),
the first interval selects the consecutive residues
\[
\varrho+i-K+2,\ldots,\varrho-1
\pmod K.
\]
These residues are distinct and disjoint from those selected by
the second interval.

Finally, for \(i=N-1\), the starter list is
\(0,\ N-1,\ldots,N-K+1
\).
Among these starters, those satisfying \(j\geq\varrho\) are
\(N-1,\ldots,N-K+1\). They select every coordinate except
\(\varrho\), exactly once.

Thus, in every case, the rows of \(\mathbf Q_i\) contributed by
starter nodes \(j\geq\varrho\) are pairwise distinct standard basis
vectors.

Let \(S_i\) be the set of starter nodes in the list that belong to
\(\{0,\ldots,\varrho-1\}\). Let \(C_i\) be the set of coordinates
not selected by the rows contributed by starters \(j\geq\varrho\).
The \(K-|S_i|\) rows contributed by starters \(j\geq\varrho\)
select exactly \(K-|S_i|\) distinct coordinates. Hence, the number
of remaining coordinates is
\[
|C_i|=K-(K-|S_i|)=|S_i|.
\]

We now permute the rows and columns of \(\mathbf Q_i\). First, place
the \(K-|S_i|\) rows contributed by starters \(j\geq\varrho\) at
the top. These rows are distinct standard basis vectors. Next, place
their selected columns on the left, in the order matching these rows.
The top-left block is then \(\mathbf I_{K-|S_i|}\), and the
top-right block is zero.
The remaining rows are those of \(\mathbf W\) indexed by \(S_i\),
and the remaining columns are indexed by \(C_i\). Writing
\(C_i^c=\{0,\ldots,K-1\}\setminus C_i
\), 
the permuted matrix has the block form
\[
\begin{bmatrix}
\mathbf I_{K-|S_i|} & \mathbf 0\\
\mathbf W[S_i,C_i^c] & \mathbf W[S_i,C_i]
\end{bmatrix}.
\]
Here, \(\mathbf W[S_i,C_i]\) denotes the submatrix with rows
indexed by \(S_i\) and columns indexed by \(C_i\). The row and
column orders in each submatrix are those induced by the
permutations above.

This matrix is block lower triangular, so its determinant is
\[
\det(\mathbf I_{K-|S_i|})
\det\bigl(\mathbf W[S_i,C_i]\bigr)
=
\det\bigl(\mathbf W[S_i,C_i]\bigr).
\]
Since row and column permutations change the determinant only by a
sign,
\[
\det(\mathbf Q_i)
=
\pm\det\bigl(\mathbf W[S_i,C_i]\bigr).
\]

We now identify \(\mathbf W[S_i,C_i]\) in each case. Coordinate
indices in the following lists are taken modulo \(K\), and empty
index ranges are omitted.
\begin{itemize}
    \item For \(0\leq i\leq\varrho-2\),
    \[
    S_i=\{0,\ldots,i+1\},
    \qquad
    C_i=\{\varrho,\ldots,\varrho+i+1\}.
    \]

    \item For \(\varrho-1\leq i\leq K-2\),
    \[
    S_i=\{0,\ldots,\varrho-1\},
    \qquad
    C_i=\{i+2,\ldots,i+\varrho+1\}.
    \]

    \item For \(K-1\leq i\leq K+\varrho-3\),
    \[
    S_i=C_i=\{i-K+2,\ldots,\varrho-1\}.
    \]

    \item For \(K+\varrho-2\leq i\leq N-2\), we have
    \(S_i=C_i=\varnothing\). All rows of \(\mathbf Q_i\) are
    distinct standard basis vectors, so \(\mathbf Q_i\) is a
    permutation matrix.

    \item For \(i=N-1\),
    \[
    S_i=\{0\},
    \qquad C_i=\{\varrho\}.
    \]
\end{itemize}

In the first three cases, Lemma~\ref{lem:determinant} ensures that
\[
\det\bigl(\mathbf W[S_i,C_i]\bigr)=\pm1,
\]
when the indices are ordered as listed. Any other ordering changes
only the sign. In the last case, the construction gives
\(\mathbf W[0,\varrho]=1\). Together with the block-triangular form above,
these observations imply
\[
\det(\mathbf Q_i)=\pm1
\qquad\text{for every }i\in\mathcal V.
\]
Thus, \(\mathbf Q_i\) is invertible over every field.
\end{proof}